\documentclass[aps,prx,amsmath,amssymb,superscriptaddress,notitlepage,twocolumn,groupedaddress]{revtex4-1}

\usepackage{comment}
\usepackage{amsmath}
\usepackage{amssymb}
\usepackage{latexsym}
\usepackage{graphicx}
\usepackage{times,psfrag}
\usepackage{enumerate}
\usepackage{amsmath}
\usepackage{dsfont}
\usepackage{dcolumn}
\usepackage{bm,bbm}
\usepackage{hyperref}
\usepackage{color}
\usepackage{latexsym,amsmath,amssymb,bm,euscript}
\usepackage{dsfont}
\usepackage{textcomp}
\usepackage{tabularx}
\usepackage{setspace}
\usepackage{sidecap}
\usepackage{placeins}
\usepackage{threeparttable}
\usepackage{multirow}
\usepackage{physics}
\usepackage{tikz-feynman}
\usepackage{makecell}
\usepackage{pifont}
\usepackage{subcaption}
\usepackage{caption}

\let \oldbm \bm
\renewcommand{\vec}[1]{\oldbm{#1}}

\def\bk{{\vec k}}
\def\CP{{\rm CP}}

\def\bn{{\vec n}}
\def\bm{{\vec m}}

\def\br{{\vec r}}

\def\Sp{\rm{Sp}}

\def\Spin{\rm{Spin}}

\def\tr{\mathop{\mathrm{tr}}}
\def\Z{\mathds{Z}}

\def\T{\mathcal{T}}
\def\L{\mathcal{L}}

\def\M{\mathcal{M}}

\def\diag{{\rm diag}}

\def\U{{\rm U}}

\def\SU{{\rm SU}}

\def\SO{{\rm SO}}
\def\O{{\rm O}}

\newcommand{\beq}{\begin{equation}}
\newcommand{\eeq}{\end{equation}}
\newcommand{\beqarray}{\begin{eqnarray}}
\newcommand{\eeqarray}{\end{eqnarray}}

\begin{document}

\title{Non-unitary pairing, fractional vortices, and charge $4e$ superconductivity in twisted bilayer graphene}

\title{Symmetry constraints on superconductivity in twisted bilayer graphene:\\ Fractional vortices, $4e$ condensates or non-unitary pairing}
\author{Eslam Khalaf}
\affiliation{Department of Physics, Harvard University, Cambridge, MA 02138}

\author{Patrick Ledwith}
\affiliation{Department of Physics, Harvard University, Cambridge, MA 02138}

\author{Ashvin Vishwanath}
\affiliation{Department of Physics, Harvard University, Cambridge, MA 02138}

\begin{abstract}
When two graphene sheets are twisted relative to each other by a small angle, enhanced correlations lead to superconductivity whose origin remains under debate. Here, we derive some general constraints on superconductivity in twisted bilayer graphene (TBG), independent of its underlying mechanism. Neglecting weak coupling between valleys, the global symmetry group of TBG consists of independent spin rotations in each valley in addition to valley charge rotations, $ \SU(2) \times \SU(2) \times \U_V(1) $. This symmetry is further enhanced to a full $\SU(4)$ in the idealized chiral limit. In both cases, we show that any charge $2e$ pairing must break the global symmetry. Additionally, if the pairing is unitary the   resulting superconductor admits {\em fractional} vortices. This leads to the following general statement: Any superconducting condensate in either symmetry class has to satisfy one of three possibilities: (i) the superconducting pairing is non-unitary, (ii) the superconducting condensate has charge $2e$ but admits at least half quantum vortices which carry a flux of $h/4e$, or (iii) the superconducting condensate has charge $2me$, $m>1$, with vortices carrying $h/2me$ flux. The latter possibility can be realized by a symmetric charge $4e$ superconductor ($m=2$). Non-unitary pairing (i) is expected for superconductors observed in the vicinity of flavor polarized states. On the other hand, in the absence of flavor polarization, e.g. in the vicinity of charge neutrality, one of the two exotic possibilities (ii) and (iii) is expected. We sketch how all three scenarios can be realized in different limits within a strong coupling theory of superconductivity based on skyrmions. Finally we discuss the effect of symmetry lowering  anisotropies and experimental implications of these scenarios.
\end{abstract}

\maketitle

\emph{\bf Introduction}--- The discovery of superconductivity in magic angle twisted bilayer graphene (TBG) \cite{PabloSC} has inspired immense research activity to determine its origin \cite{Dean-Young, Efetov, EfetovScreening, YoungScreening, Xu2018, phononLianBernevig, phononMacDonald, YouVishwanath, IsobeFuSC, ChubukovVanHove, ChubukovTBG, FernandesVanderbos, KangFernandes, Kozii, SkPaper, SkDMRG}. Several mechanisms for the superconductivity were proposed including phonon-induced pairing \cite{phononLianBernevig, phononMacDonald}, bosonic order parameter fluctuations \cite{YouVishwanath, IsobeFuSC, ChubukovVanHove, ChubukovTBG, FernandesVanderbos, KangFernandes, Kozii}, and topological skyrmion superconductivity \cite{SkPaper, SkDMRG, Christos}. At the same time, a few works have attempted to connect band topology to superfluid density \cite{BernevigTBGTopology, TBGV} or categorized different paired states on the basis of symmetry \cite{ScheurerSymm}. TBG at small twist angles has a rather large symmetry, particularly when we ignore weak Hunds coupling that couples spins in opposite valleys. Here, we exploit this large symmetry group to expose   strong constraints on the nature of superconductivity regardless of the mechanism responsible for it. Remarkably, these constraints imply that the superconductivity has to be exotic in one of three possible ways: either (i) the pairing is non-unitary \cite{Leggett75, Volovik87} which means that the different flavors pair independently, (ii) the pairing breaks the global symmetries leading to an order parameter manifold which admits fractional vortices, (iii) the superconducting condensate has charge $2me$ with $m>1$ admitting fractional vortices with flux $h/2me$. Each of these scenarios leads to measurable physical consequences. 

A notable feature of magic angle TBG is that electrons appear with both spin and valley flavors. The absence of spin-orbit coupling in graphene makes $\SU(2)_S$ spin rotation an excellent symmetry. Similarly, the small angle leads to an additional $\U(1)_V$ symmetry,  implying conservation of the valley quantum number, in addition to charge $\U(1)_C$ conservation. Finally, again in the small angle limit, the independent spin rotation symmetry in the two valleys is only weakly broken by a Hunds coupling, which can be ignored for many purposes. Thus while the low energy symmetry group is $\SU(2)_S \times \U(1)_C \times \U(1)_V$, it is  meaningful to also consider the symmetry group $\U(2)_K \times \U(2)_{K'}$ denoting independent $\U(2)$ symmetry groups in the opposite valleys. Finally, several works \cite{KIVCpaper, KangVafekPRL, Uchoa, VafekKangRG} have suggested that this symmetry is enhanced to a full $ \U(4)$ flavor symmetry to a good approximation. This approximation becomes exact in the so-called chiral limit \cite{Tarnopolsky}. In the following, we will omit the overall charge conservation $\U(1)_C$ which will always be assumed and denote the three possible symmetry settings by $G_{\rm low} = \U_V(1) \times \SU(2)_S$, $G = \U_V(1) \times \SU(2)_K \times \SU(2)_{K'}$, and $G_{\rm high} = \SU(4)$. A hierarchy of energy scales lowers the symmetry in steps $G_{\rm high} \rightarrow G \rightarrow G_{\rm low}$ \cite{KIVCpaper}. For the most part, we will ignore the weak intervalley Hunds coupling and concern ourselves with either $G$ or the enlarged symmetry $G_{\rm high}$. At the  end we will discuss  how the conclusions are modified on including the weak Hunds coupling, which can be either ferromagnetic or antiferromagnetic, and close with a discussion about the relevance to experiment.\\

\emph{\bf Summary of the results}--- Before presenting the details of our analysis, let us summarize our main results. First, we note that the pairing function is a matrix in the space of flat bands labelled by band, spin, and valley  indices. When considering pairing in a multi-orbital superconductor, it is important to distinguish the cases of unitary and non-unitary pairing first discussed in the context of Helium 3 \cite{Leggett75, Volovik87}. Unitary pairing corresponds to the case where the pairing matrix is proportional to a unitary matrix which guarantees that its eigenvalues has the same magnitude leading to a single pairing gap. In non-unitary pairing, this is not the case and we can get different pairing gaps for different flavors e.g. different gaps for the two spin species. Usually, unitary pairing is energetically favored when the normal state neither breaks flavor symmetry \cite{UCoGe, Uranium} nor is close to a flavor-symmetry breaking phase transition \cite{Nevidomskyy, UTeScience}. 

Our result can be phrased as a symmetry imposed constraint on the type of unitary pairing possible in TBG. This follows from the simple observation that there is no symmetric charge $2e$ superconductor either under $G_{\rm high} = \SU(4)$ or $G= \U_V(1) \times \SU(2)_K \times \SU(2)_{K'}$ leaving one of two possibilities. First, a charge $2e$ superconductor which spontaneously breaks the global symmetry ($G$ or $G_{\rm high}$) by transforming as a non-trivial representation. As we will show, such superconductors always admit exotic vortices which carry a fractional flux. These arise from the non-trivial flavor structure of the symmetry breaking pairing function which allows us to compensate for a rotation in the phase of the condensate with an internal rotation in the flavor space. This is in complete analogy with the emergence of half-vortices in liquid Helium 3 or spin-triplet superconductors \cite{Volovik87, HQVHe3, SrRuO} where a $\pi$ phase rotation of the overall phase is combined with an internal rotation in spin space.  The second possibility is a ($G$ or $G_{\rm high}$) charge $2me$ superconductor with $m>1$. Such higher charge superconductor is allowed to be symmetric under the global symmetry and it will automatically admit fractional vortices with flux $h/2me$. Higher charge superconductivity can be easily detected experimentally through the periodicity of the Fraunhofer pattern in Josephson junctions. In this regard, we note an important distinction between the fractional vortices  of the $2e$  superconductor and  the fractional vortices in higher charge symmetric superconductors. While the former have logarithmically divergent energies, even in the presence of a magnetic field,  and can only be observed in certain mesoscopic geometries \cite{Babaev2001, Silaev, Kim2007}, the latter has a finite energy and can be stabilized in the infinite system. This difference can be understood by recalling the difference between vortices in a superconductor whose logarithmic divergence is cancelled by the coupling to the electromagnetic gauge field and a neutral superfluid where the logarithmic divergence persists. The logarithmically divergent energy of the $2e$ fractional vortices arise precisely because they combine a rotation of the superconducting phase, which can be cancelled by an external flux, with an internal flavor rotation which cannot. On the other hand, fractional vortices in symmetric higher charge superconductors arise from pure phase winding which can be completely cancelled with an external flux.

\begin{figure}
    \centering
    \includegraphics[width = 0.5 \textwidth]{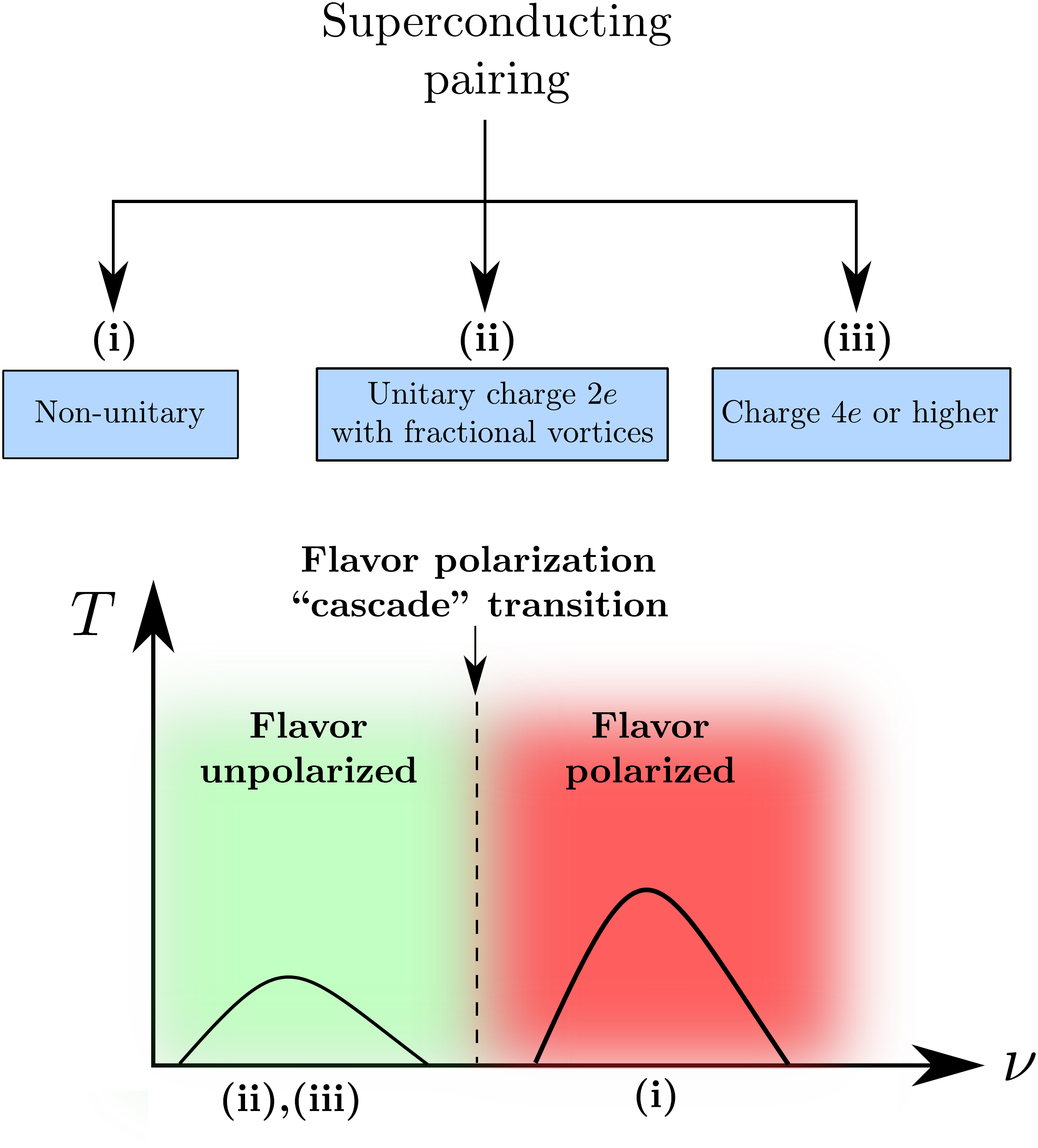}
    \caption{Schematic illustration of the three possibilities for superconducting pairing with the global symmetry $G = \U(1)_V \times \SU(2)_K \times \SU(2)_{K'}$ or $G_{\rm high} = \SU(4)$. Non-unitary pairing (i) is expected in TBG for superconductivity in the vicinity of flavor symmetry breaking whereas unitary charge $2e$ superconductivity with fractional vortices (ii) or higher charge superconductivity (iii) is expected in the {\em absence} of flavor symmetry breaking. The unitary charge $2e$ superconductor always breaks the global ($G$ or $G_{\rm high}$) symmetry and is thus only possible at $T = 0$. The possible order parameter manifolds for the unitary charge $2e$ superconductor and its associated vortices are summarized in Table \ref{SC2e}.}
    \label{fig:schematic}
\end{figure}

We  discuss the implications of these results for finite temperature physics and point out how, in certain cases, $2e$-superconductors with fractional vortices are reduced by thermal fluctuations to higher charge superconductors.  We also demonstrate how these different scenarios for superconductivity are explicitly realized in the recently proposed theory of skyrmion superconductivity \cite{SkPaper, SkDMRG}, by constructing both $2e$ and $4e$ charged skyrmions near charge neutrality. We note here a recent work \cite{Christos} which also investigated topological terms in TBG within a Dirac approximation and found level 1 WZW terms for half-filling and level 2 WZW term for neutrality. These correspond to cases (i) and (iii) in our analysis (see Fig.~\ref{fig:schematic}). Finally, we discuss the effects of further symmetry lowering perturbations such as the intervalley Hunds coupling.

To discuss the implications of these results for TBG, we need to distinguish the cases of pairing in the presence or absence of flavor polarization. This distinction can be made more clear by recalling some basic facts regarding superconductivity in the presence of ferromagnetic or antiferromagnetic order. Although both spontaneously break $\SU(2)$ rotation symmetry, only the former has finite spin polarization, i.e. the state has a finite charge under a $\U(1)$ subgroup of $\SU(2)$. This means that, at least in weak coupling, we have a spin split Fermi surface where we expect pairing to take place independently leading generically to non-unitary pairing. On the other hand, an antiferromagnet breaks $\SU(2)$ but still retains some discrete symmetries relating the two spin species which usually results in the same pairing amplitude in the two spin species, leading generically to unitary pairing. These ideas can be generalized to TBG by definition a flavor polarized state to be any state which carries a finite charge under any $\U(1)$ subgroup of $G$, indicating finite spin or valley polarization. Note that many symmetry breaking state are not flavor polarized in analogy to the antiferromagnetic case. For instance, the intervalley coherent states considered in Ref.~\cite{KIVCpaper} or the $C_2 \T$-breaking valley Hall states \cite{ShangHF, MacdonaldHF, KIVCpaper} are \emph{not} flavor polarized despite breaking some symmetries since they still manage to retain some discrete symmetries (the so-called Kramers time-reversal symmetry \cite{KIVCpaper}).

This leads to the following conjectures about the nature of superconducting pairing in TBG, summarized in Fig.~\ref{fig:schematic}. First, the superconductors obtained in regions with flavor polarization -- which can be identified experimentally through the evolution of the Hall number or quantum oscillations -- likely correspond to non-unitary pairing, which is not constrained by our theory. This corresponds to the majority of the superconductivity observed in experiments close to half filling \cite{PabloSC, PabloMott, Dean-Young, YoungScreening, CaltechSC}. However, in certain isolated cases, superconductivity in the vicinity of charge neutrality \cite{Efetov, EfetovScreening}, or reached from charge neutrality in the absence of flavor polarization \cite{EfetovScreening, CaltechSC, PabloTrilayer}, has also been observed. We conjecture that these cases present a fertile hunting ground for exotic forms of superconductivity, either paired superconductors featuring fractional vortices or even higher charge condensates such as $4e$ superconductors. We hope future experiments will search for signatures of such novel condensates.\\

\emph{\bf Symmetry transformation of charge $2e$ superconductors}--- 
Let us begin by considering the transformation properties of an arbitrary charge $2e$ superconductor under a general unitary transformation. A charge $2e$ superconductor is generally described by an order parameter $\Delta$ which transforms as $\Delta \mapsto \Delta e^{2i \phi}$ under global $\U_C(1)$ charge conjugation which acts on the electrons as $c \mapsto e^{i \phi} c$. This breaks $\U_C(1)$ down to $\Z_2$. We take the electron operators $c_a$ to be labelled by an index $a$. For example, in a translationally symmetric system, this index $a = (\alpha, \bk)$ includes both orbital $\alpha$ and momentum $\bk$ degrees of freedom.  The pairing function is thus taken to be a matrix of the form $\Delta_{ab} = \langle c_a c_b \rangle$. Under a general unitary transformation $c \mapsto U c$, $\Delta$ transforms the same way as $c c^T$ leading to
\beq
\Delta \mapsto U \Delta U^T
\label{UDelta}
\eeq
Note that this relation holds as long as $\Delta$ transforms the same way as $c c^T$ and does not make any addition assumption on its nature. In particular, it holds well beyond the weak coupling BCS limit.

Throughout most of this paper, we will focus on global symmetries which only act on the orbital index $\alpha$ of the electron operator. This orbital index will be taken to label the 8 flat bands of TBG. We employ the Chern-sublattice-spin basis defined in Refs.~\cite{KIVCpaper, SkPaper} with Pauli matrices $\gamma_z = \pm$ distinguishing the $\pm$ Chern sectors, $\eta_z = \pm = K/K'$ distinguishing the two valleys and $s_z = \pm = \uparrow/\downarrow$ distinguishing the two spins. The generators of $\U(4) = \U(1)_C \times G_{\rm high}$ are $\gamma_0 \eta_\mu s_\nu$ where $\mu, \nu = 0,x,y, z$ whereas the generators of $\U(2) \times \U(2) = \U(1)_C \times G$ are $\gamma_0 \eta_{0,z} s_\nu$. 

We first note that the generators for $G_{\rm high}$ (and as a result also the generators of $G \subset G_{\rm high}$) are all proportional to $\gamma_0$. This means that the pairing function can be decomposed into different pairing channels in the Chern space which can be considered independently
\beq
\Delta = \sum_{\mu=0,x,y,z} \Delta_\mu \gamma_\mu,
\eeq
Here, $\Delta_\mu$ satisfies
\beq
\Delta_\mu^T = \begin{cases}
-\Delta_\mu &: \mu = 0,x,z \\
\Delta_\mu &: \mu = y 
\end{cases}
\label{Deltamu}
\eeq

Although the results in the remainder of this paper hold for arbitrary pairing functions $\Delta_\mu$, it is instructive to consider the case of translationally symmetric pairing between time-reversal related momenta $\bk$ and $\bk' = -\bk$ given by $\Delta(\bk,\bk') = \Delta(\bk) \delta_{\bk,-\bk'}$. This allows us to classify the pairing function according to their transformation properties under spatial rotations into $s$, $p$, $d$, etc. We then see from Eq.~\ref{Deltamu} that $\Delta_{0,x,z}(-\bk)^T = -\Delta_{0,x,z}(\bk)$ whereas $\Delta_y(-\bk)^T = \Delta_y(\bk)$. Thus, antisymmetric pairing in orbital space $\Delta_\mu^T(\bk) = -\Delta_\mu(\bk)$ is associated with even momentum pairing ($s$, $d$, etc) for $\Delta_{0,x,z}(\bk)$ and with odd momentum pairing ($p$, $f$, etc) for $\Delta_y(\bk)$. For symmetric pairing, the opposite is true. Physically, we expect the pairing between Chern sectors $\Delta_{x,y}$ to be favored compared to pairing within a Chern sector $\Delta_{0,z}$. However, in the following, we will keep our discussion as general as possible. 

Let us briefly comment on the role of spatial symmetries. First, they place constraints on the $\bk$-dependence of the pairing function $\Delta_\mu(\bk)$. In addition, as shown in the supplemental material, symmetries which exchange valleys, such as mirror or two-fold rotation,  relate the two intra-Chern pairing channels $\gamma_\pm = \frac{1 \pm \gamma_z}{2}$ which will then form a higher-dimensional representation of the full symmetry group. These considerations, which are analyzed in detail in the supplemental material, do not affect any of the conclusions we will derive below which only rely on the transformation properties under global symmetries. For notational simplicity, we will usually drop the $\bk$ dependence of the pairing function since we are mostly concerned with the symmetry transformation properties under global symmteries which only act on the internal (orbital) indices. We will also drop the $\mu$ subscript since the action of the global symmetries do not mix the different Chern channels.\\

\emph{\bf Absence of symmetric charge $2e$ superconductors}--- Let us first consider the high-symmetry limit with $G_{\rm high} = \SU(4)$ whose action on $\Delta$ is given by Eq.~\ref{UDelta}. One fundamental observation is that it is not possible to choose a non-zero pairing matrix $\Delta$ which transforms as a singlet under $\SU(4)$. Instead, the action of $\SU(4)$ on $\Delta$ decomposes into a sum of a 6-dimensional and a 10-dimensional representations. This can be understood by noting that the action of $\SU(4)$ corresponds to the tensor product of two copies of the fundamental representation $\boldsymbol 4$ which decomposes as ${\boldsymbol 4} \otimes {\boldsymbol 4} = {\boldsymbol 6} \oplus \boldsymbol{10}$. The ${\boldsymbol 6}$ and $\boldsymbol{10}$ correspond to the antisymmetric and symmetric tensor representations of $\SU(4)$, respectively. 

Notice that $\SU(4)$ symmetry mixes inter-valley and intra-valley pairing and there is no sense in distinguishing the two. This may seem at odds with our physical expectation that pairing takes places between time-reversal related states which live in opposite valleys. However, note that $\SU(4)$ allows for unitary rotations which mix the two valleys. As a result, we can define a modified time-reversal symmetry $\T'$ that acts within the same valley by combining such rotation with the standard time-reversal $\T$. The existence of intra-valley time-reversal was noted in Ref.~\cite{KIVCpaper} and more recently studied in Ref.~\cite{CanoChiral}.

Next, we can consider what happens in the more realistic limit where $G_{\rm high}$ is broken down to $G = \U(1) \times \SU(2) \times \SU(2)$. Let us first consider the  $\SU(2) \times \SU(2)$ part. The antisymmetric representation $\boldsymbol{6}$ splits into $\boldsymbol{4} \oplus \boldsymbol{1} \oplus \boldsymbol{1}$ under $\SU(2) \times \SU(2)$ whereas the symmetric representation splits into $\boldsymbol{10} = \boldsymbol{4} \oplus \boldsymbol{3} \oplus \boldsymbol{3}$. These can be understood as follows: first recall the representations of $\SU(2)$ are labelled by a half-integer $S$ with dimension $2S + 1$. Thus, the 1D irrep of $\SU(2) \times \SU(2)$ is simply the singlet representation in both valleys $(S_K, S_{K'}) = (0, 0)$. The 3D irreps above correspond to a singlet in one valley and triplet in the other $(S_K, S_{K'}) = (0,1)$ or $(1, 0)$ whereas the 4D irreps correspond to $S = 1/2$ in both valleys. Alternatively, this 4D irrep can be understood as the fundamental of $\Spin(4)$ which is isomorphic to $\SU(2) \times \SU(2)$.  Note that, unlike the $\SU(4)$ case, it is possible to have a $\SU(2) \times \SU(2)$ singlet. These correspond to the intravelley pairing channels $\propto \eta_{0,z}$. However, these pairing channels carry a charge under $\U_V(1)$. Thus, similar to the $G_{\rm high}$ case, there is no symmetric pairing channel under the full $G$ symmetry. We note that our statement for the symmetry group $G$ is consistent with the symmetry analysis of Ref.~\cite{ScheurerSymm}.

Although the above argument applies to all pairing channels, the most natural expectation in the case of the symmetry group $G$ is pairing between time-reversal related electrons in opposite valleys  \cite{YouVishwanath, IsobeFuSC, ChubukovVanHove, ChubukovTBG}. Restriction to intervalley pairing is equivalent to the condition $\{ \Delta, \eta_z \} = 0$. This selects 8 of the 16 generators of $\U(4)$  given explicitly by $\eta_{x,y} s_\nu$, $\nu = 0,x,y,z$. Among these generators, four are symmetric and four are antisymmetric. Thus, the action of $\SU(2) \times \SU(2)$ on the inter-valley pairing (which is invariant under $\U_V(1)$) splits into $\boldsymbol{4} \oplus \boldsymbol{4}$. For the remainder of this work, we will focus on the physically relevant case of intervalley pairing. The case of intravalley pairing is discussed in the supplemental material for completeness.\\

 \emph{\bf Charge $4e$ superconductors}--- The previous discussion leaves open the more intriguing possibility of a symmetric charge $4e$ condensate. The existence of such condensate can be seen by first writing the charge $4e$ pairing function
 \beq
 \Delta_{\alpha \beta \gamma \delta} = \langle c_{\alpha} c_{\beta} c_{\gamma} c_{\delta} \rangle
 \eeq
 $ \Delta_{\alpha \beta \gamma \delta}$ transforms as
 \beq
  \Delta_{\alpha \beta \gamma \delta} \mapsto U_{\alpha \alpha'} U_{\beta \beta'} U_{\gamma \gamma'} U_{\delta \delta'} \Delta_{\alpha' \beta' \gamma' \delta'}
 \eeq
 A singlet pairing channel is obtained by taking $\Delta$ to be proportional to antisymmetric tensor $\Delta_{\alpha \beta \gamma \delta} \propto \epsilon_{\alpha \beta \gamma \delta}$ symbol since
 \beq
 U_{\alpha \alpha'} U_{\beta \beta'} U_{\gamma \gamma'} U_{\delta \delta'} \epsilon_{\alpha' \beta' \gamma' \delta'} = \det U \epsilon_{\alpha \beta \gamma \delta} = \epsilon_{\alpha \beta \gamma \delta}
 \eeq
 Thus, in contrast to the charge $2e$ superconductor, an $\SU(4) = G_{\rm high}$ symmetric charge $4e$ superconductor is possible. This conclusion also holds for the symmetry group $G \subset G_{\rm high}$. \\
 
 \emph{\bf Unitary pairing and fractional vortices}---
 We start by discussing the case of unitary pairing \cite{Leggett75} which is typically energetically favored if pairing takes place equally in all flavors. In TBG, we expect unitary pairing in the vicinity of charge neutrality or in samples with no signature of flavor symmetry breaking \cite{Efetov, EfetovScreening} as will be discussed later.  For unitary pairing, $\Delta \Delta^\dagger$ is proportional to the identity matrix so that we can rescale $\Delta$ to be unitary (away from the points where it vanishes).  
 
 To study different types of vortices, we start by noting that the overall phase for a unitary superconductor described by an $n \times n$ matrix order parameter $\Delta$ is defined as
 \beq
 \varphi = \frac{1}{n} \arg \det \Delta
 \label{phi}
 \eeq
 This definition guarantees that the phase changes as $\varphi \mapsto \varphi + \phi$ whenever we multiply $\Delta$ by an overall phase $\phi$. The phase $\varphi$ couples to the background gauge field with charge $2e$ according to the Ginzburg-Landau (GL) functional
 \beq
 \L \propto (\hbar \partial_\mu \varphi - 2 e A_\mu)^2 + \dots
 \label{GL}
 \eeq
 Variations with respect to $A$ leads to
 \beq
\int d l_\mu A_\mu = \frac{\hbar}{2 e} \int d l_\mu \partial_\mu \varphi
\label{Flux}
 \eeq
 which is equivalent to the condition of vanishing current in the ground state. Let us see the implications of this for the different symmetry settings.
 
 Let us first discuss the more physical $G$ symmetry and focus on the physically relevant intervalley pairing. This can be decomposed into a symmetric and an antisymmetric component given by
 \beq
 \Delta^{\rm S/A} =  \left(\begin{array}{cc}
    0 & \tilde \Delta \\
    \pm \tilde \Delta^T & 0
 \end{array}\right)_\eta
 \label{DeltaInterValley}
 \eeq
 where $\tilde \Delta$ is unitary. Notice that due to the $\SU(2) \times \SU(2)$ symmetry, the $\SU(2)$ spin singlet and triplet components are joined in a single 4-dimensional irrep. As a consequence, we cannot split $\Delta^{\rm S/A}$ further into a valley-singlet spin-triplet channel and a valley-triplet spin-singlet channel \cite{TDBGTheory, ScheurerSymm}. This is only possible in the presence of Hund's coupling which breaks $\SU(2)_K \times \SU(2)_{K'}$ down to $\SU(2)_S$ as we will discuss later. For either the symmetric or antisymmetric representation, we can write $\varphi = \frac{1}{4} \arg \det \Delta = \frac{1}{2} \arg \det \tilde \Delta$. Using (\ref{Flux}), we see that if only one of the two eigenvalues of $\tilde \Delta$ winds by $2\pi$, the phase $\varphi$ winds by $\pi$ leading to a vortex with a flux of $h/4e$.
 
  The case of $\SU(4)$ symmetry can be analyzed similarly. For the antisymmetric representation with $\Delta^T = -\Delta$, the eigenvalues come in pairs $\{\lambda, -\lambda\}$ which means that the winding of $\varphi$ is necessarily a multiple of $\pi$ leading also to half-vortices. For the symmetric representation, there is no constraint on the winding of individual eigenvalues of $\Delta$, which means that the winding of $\varphi$ is a multiple of $\pi/2$ leading to \textit{quarter} vortices with flux $h/8e$ in this case.
  
   Finally note that the case of intravalley pairing for the symmetry group $G$ can be obtained from the high symmetry case $G_{\rm high}$ as follows. Unitary intravalley pairing splits into an antisymmetric representation corresponding to spin-singlet in both valleys and a symmetric representation corresponding to spin-triplet pairing in both valleys. These can be obtained by restricting the $\SU(4)$ antisymmetric and symmetric representations using the condition $[\Delta, \eta_z] = 0$ which does not affect the flux quantization condition leading to $h/4e$ vortices for the former and $h/8e$ vortices in the latter. A more detailed analysis of these cases is given in the supplemental material.\\
   
   \emph{\bf Order parameter manifolds for unitary pairing}--- We will now present an alternative and more abstract argument for the existence of fractional vortices in the different unitary charge $2e$ superconductors. This argument will not rely on any GL functional. Let us again start with the case of intervalley pairing with $G$ symmetry. The symmetric/antisymmetric pairing channels can be generated starting with an arbitrary symmetric/ antisymmetric valley off-diagonal matrix and applying $\U(2) \times \U(2)$ symmetry according to (\ref{UDelta}). The resulting pairing matrix is
 \beq
 \Delta^{\rm S,A} = \left(\begin{array}{cc}
    0 & U_1 U_2^T \\
    \pm U_2 U_1^T & 0
 \end{array}\right)_\eta, \quad \implies \quad \tilde \Delta = U_1 U_2^T
 \eeq
 Thus, $\tilde \Delta$ is parametrized by a pair of $2 \times 2$ unitary matrices $U_1$ and $U_2$ modulo the operation $U_1 \mapsto U_1 V$, $U_2 \mapsto U_2 V^*$ for any $2 \times 2$ unitary matrix $V$. This means that $\tilde \Delta$ parametrizes the coset space $\U(2) \times \U(2) / \U(2) \simeq \U(2)$. The existence of half-vortices can be understood as follows. A full vortex is a vortex in the overall phase of $\tilde \Delta$ which corresponds to a loop in the subgroup $\U_C(1)$ generated by $\tilde \Delta = e^{i \phi}$ with $\phi: 0 \rightarrow 2\pi$. On the other hand, we can write the decomposition $\U(2) = \frac{\U_C(1) \times \SU(2)}{\Z_2}$ which is seen more explicitly by writing $\tilde \Delta$ as $e^{i \varphi} V$ with $V \in \SU(2)$. The identification of $(\varphi, V)$ and $(\varphi + \pi, -V)$ allows for a $\pi$ vortex in the superconducting phase $\varphi$ combined with a rotation which sends $V$ to $-V$. This can be written as $\pi_1(\U(2)) = \Z/2$, where we identify the element $1 \in \Z$ with the fundamental loop in $\U_C(1)$ generated by $\tilde \Delta = e^{i \phi}$, $\phi: 0 \rightarrow 2\pi$. The minimal loop in this space $1/2 \in \Z/2$ thus corresponds to a half quantum vortex with $h/4e$ flux.

 The existence of half (quarter) vortices for the antisymmetric (symmetric) repsenetations of $\SU(4)$ can also be understood by analyzing the order parameter manifolds. For the antisymmetric representation $\Delta_\mu^A$, the manifold is generated by acting with $\U(4)$ on any given unitary antisymmetric matrix according to (\ref{UDelta}) leading to
\beq
\Delta^A = U \eta_y s_0 U^T
\eeq
where $U \in \U(4)$. We notice that $\Delta_\mu^A$ is invariant under multiplying $U$ from the right by any unitary matrix $K$ satisfying
\beq
K \eta_y s_0 K^T = \eta_y s_0
\eeq
 which is the definition of a $4 \times 4$ symplectic matrix $K \in \Sp(4)$. Thus, $\Delta_\mu^A$ parametrizes the coset space $\U(4)/\Sp(4)$. Writing $U = e^{i \theta} V$ where $V \in \SU(4)$, we see that we can form a closed path in $\U(4)$ by taking $V \mapsto -i V$ and $\theta \mapsto \theta + \pi/2$. We note however that doing this twice yields a regular vortex since the superconducting phase $\varphi$ is twice $\theta$ and the path $V \mapsto -V$ can be deformed into a path $K \mapsto -K$ lying completely in the symplectic group which is identified with the identity. Thus, $\pi_1\left(\frac{\U(4)}{\Sp(4)}\right) = \Z/2$. A similar argument applies to the symmetric 10-dimensional representation. In this case, the order parameter manifold $\frac{\U(4)}{\O(4)}$ which admits half-vortices similar to the case of $\frac{\U(4)}{\Sp(4)}$. However, due to the structure of the $\O(4)$ group, we can also form \emph{quarter} vortices. These are obtained by taking a trajectory in $\U(4)$ which starts at an element $K_+ \in O(4)$ with $\det K_+ = 1$ and ends at an element $K_-$ with $\det K_- = -1$. Since $O(4) \subset U(4)$ and $\U(4)$ is path connected, we can always find such trajectory. This trajectory is a loop in the coset space $\frac{\U(4)}{\O(4)}$ which involves a winding of the phase of $\theta$ by $(2l+1)\frac{\pi}{4}$ yielding a quarter vortex for $l=0$ of the superconducting phase $\varphi = 2 \theta$. An explicit example of such trajectory is given by the diagonal matrix $U = \diag(1, 1, 1, e^{i \theta})$, $\theta: 0 \mapsto \pi$ which connects the identity to the matrix $U = \diag(1,1,1,-1)$ which is contained in $\O(4)$.
 
In general, we can derive the flux quantization condition without a GL function by employing the following two axioms: (a) the flux is identified as a homomorphism from the first homotopy group of the order parameter manifold, which we will denote by $\M_\Delta$, to the real numbers $F: \pi_1(\M_\Delta) \mapsto \mathbbm{R}$ and (b) the element of $\pi_1(\M_\Delta)$ corresponding to the fundamental loop in $\U_C(1)$ generated by $e^{i \phi}$, $\phi:0\rightarrow 2\pi$, is assigned a flux of $h/2e$. Using this definition, we can reproduce the half and quarter vortices identified above. For example, the full vortices of the manifold $\U(2)$ were identified with the element $1 \in \pi_1(\U(2))$ and the half vortices were identified with the element $1/2$. Since the flux $F$ is a homomorphism, we have $2 F(1/2) = F(1) = h/2e \, \implies \, F(1/2) = h/4e$. \\

 \emph{\bf Non-unitary pairing}--- Let us now see what happens if we lift the restriction of unitary pairing. Non-unitary pairing is expected if the parent state has broken flavor symmetry such that pairing takes place in each flavor separately. Although not very common, there are a few known examples for non-unitary pairing, particularly in ferromagnetic superconductors \cite{Uranium, UCoGe, UTe, FerroSC} where pairing develops on top of a spin-split Fermi surface and has different amplitude for the different spin species. Another example is the $A_1$ phase of He${}_3$ \cite{Volovik87}. In the following, we will show that for non-unitary pairing, it is possible to have no fractional vortices. For this purpose, it suffices to focus on the most physically relevant case of intervalley pairing with $G = \U(1) \times \SU(2) \times \SU(2)$ symmetry. In general, we can specify the pairing function $\Delta$ using the eigenvalues of $\Delta^\dagger \Delta$ which are invariant under the symmetry transformation (\ref{UDelta}). For inter-valley pairing, the pairing function has the form given in Eq.~\ref{DeltaInterValley}. Thus, the most general form of the pairing functions $\tilde \Delta$ is
 \beq
\tilde \Delta = U_1 \left(\begin{array}{cc}
    \lambda_1 & 0 \\
    0 & \lambda_2
 \end{array}\right) U_2^T
 \eeq
 For $|\lambda_1| \neq |\lambda_2|$, this corresponds to so-called two gap superconductors with different amplitudes of the gap function for the different electron species \cite{Babaev2001}. The GL theory in this case is given by
 \beq
 \L = \rho_1 (\partial_\mu \varphi_1 - \frac{2 e}{\hbar} A_\mu)^2 + \rho_2 ( \partial_\mu \varphi_2 - \frac{2 e}{\hbar} A_\mu)^2 + \dots
 \eeq
 where the two $\U(1)$ phases $\varphi_{1,2} = \arg \lambda_{1,2}$ correspond to the gap functions of the two different spin species.
 The vortex quantization condition is
 \beq
\int d l_\mu A_\mu = \frac{ \hbar}{2 e} \int d l_\mu \frac{\rho_1 \partial_\mu \varphi_1 + \rho_2 \partial_\mu \varphi_2}{\rho_1 + \rho_2}
\label{Flux12}
 \eeq
 We see that this generically leads to fractional vortices which depend on the relative stiffness of the two superconductors when both $\lambda_1$ and $\lambda_2$ are non-zero \cite{Babaev2001}.
 
 The existence of fractional vortices can also be understood from our more abstract definition in the previous section. The order parameter manifold can be understood as follows. Allowing $\lambda_1$ and $\lambda_2$ to be complex, we can absorb the overall phases of $U_1$ and $U_2$ as well as the effect of the transformations $U_{1,2} \mapsto e^{i \phi_{1,2} \sigma_z} U_{1,2}$ in the phases of $\lambda_1$ and $\lambda_2$. This means that $\tilde \Delta$ parametrizes the manifold $\U(1) \times \U(1) \times S^2 \times S^2$ which can be understood as follows. Due to the independent $\SU(2)$ spin rotation in each valley, we can choose the spin quantization axis independently in the two valleys leading to two unit vectors parametrizing $S^2 \times S^2$. The two $\U(1)$ phases describe the phase of the pairing function for the two spin species. Note that $\pi_1(\U(1) \times \U(1) \times S^2 \times S^2) = \Z \oplus \Z$ where the fundemantal vortex in $\U_C(1)$ corresponds to the element $(1,1) \in \Z \oplus \Z$ which represents $2\pi$ phase winding in both $\U(1)$ phases. The flux quantization condition can then be written as $F[(1,1)] = h/2e = F[(0,1)] + F[(1,0)]$. In the absence of any symmetry relating $(0,1)$ and $(1,0)$, i.e. the two spin species, we can write $F[(0,1)] = x h/2e$ for some arbitrary $0 \leq x \leq 1$ which implies $F[(1,0)] =  (1 - x) h/2e$. Comparing with (\ref{Flux12}), we can identify $x$ with $\rho_2/(\rho_1 + \rho_2)$. Note that whenever the two spin species are related by an unbroken symmetry, $x = 1/2$ and we recover half vortices as in the unitary case.
 
In the extreme non-unitary limit where pairing takes place in only one species, i.e. one of $\lambda_{1,2}$ (and consequently $\rho_{1,2}$) is zero, there are only regular $h/2e$ vortices. This can be seen by noting that the order parameter manifold for this case reduces to $\U(1) \times S^2 \times S^2$ because phase rotations on the vanishing eigenvalue have no effect. This manifold has a single $\U(1)$ with the GL functional (\ref{GL}) leading to the same quantization condition (\ref{Flux}) wheres the winding of $\varphi$ has to be a multiple of $2\pi$ leading to standard $h/2e$ flux. This serves to illustrate that the non-unitary pairing allows for a conventional scenario with a charge $2e$ superconductor that does not host fractional vortices.\\
 
 \emph{\bf No-go theorem}--- In summary, we have proved the following general statement: {\em For any pairing in the ideal $\SU(4)$ symmetric model or the more realistic $\U(1) \times \SU(2) \times \SU(2)$ model, one of three possibilities should occur: (i) the superconducting pairing is non-unitary, (ii) the superconducting condensate has charge $2e$ but breaks the global symmetry and admits at least half quantum vortices with flux of $h/4e$ , or (iii) the superconducting condensate has charge of at least $2me$ with $m>1$ with vortices carrying $h/2me$ flux.}
 
 Note an important difference between the cases (ii) and (iii). The fractional vortices of the charge $2e$ superconductor combine a winding in the overall phase with a winding in internal orbital indices. Whereas the former is compensated by external magnetic flux, the latter is not, leading to logarithmically divergent energy at long distances. This means that these vortices are only relevant at sufficiently short distances e.g. in mesoscopic geometries. Such type of vortices were already observed in mesoscopic samples in liquid Helium 3 \cite{HQVHe3} and spin-triplet condensates \cite{HQVBEC}. On the other hand, vortices of a charge $2me$ superconductor with $m>1$ which carry a fractional flux of $h/2me$ have finite energy and can be observed in the infinite system.\\
 
 \begin{table*}[t]
     \centering
     \centering
    \bgroup
     \begin{tabular}{c|c|c|c|c}
     \hline \hline
        Symmetry  & Irreps & generators & SC Manifold (unitary pairing)  & Vortices \\
        \hline
        \multirow{2}{*}{$\SU(4)$} & $\boldsymbol{6}$ & $\eta_y s_{0,x,z}$, $\eta_{0,x,z} s_y$ &  $\frac{\U(4)}{\Sp(4)}$ & $\Z/2$ \\
        & $\boldsymbol{10}$ & $ \eta_{0,x,z} s_{0,x,z}$, $\eta_y s_y$ & $\frac{\U(4)}{\O(4)}$ & $\Z/4$ \\
        \hline
        \multirow{2}{*}{\makecell{$\SU(2) \times \SU(2) \simeq \Spin(4)$\\ Inter-valley pairing}} & $\boldsymbol{4}$ & $ \eta_y s_{0,x,z}$, $ \eta_x s_y$ & $\U(2)$ & $\Z/2$ \\
        & $\boldsymbol{4}$ & $ \eta_y s_y$, $\eta_x s_{0,x,z}$ & $\U(2)$ & $\Z/2$ \\
        \hline \hline
     \end{tabular}
     \egroup
     \caption{Summary of the order parameter manifolds and the irreducible representations for the unitary pairing charge $2e$ superconductors. For the ideal $\SU(4)$ case, the symmetry action splits into a 6-dimensional antisymmetric and a 10-dimensional symmetric irrep whose corresponding manifolds admit half and quarter quantum vortices respectively. For the more realistic $\U(1)_V \times \SU(2) \times \SU(2)$, we focus on the physically relevant inter-valley pairing which is $\U(1)_V$ singlet and splits into two 4-dimenstional irreps under $\SU(2) \times \SU(2)$ whose manifolds admit half vortices.}
     \label{SC2e}
 \end{table*}

 \emph{\bf Nonzero temperature and BKT transitions}---Here we discuss how the different scenarios we have discussed manifest at nonzero temperatures.  At nonzero temperature in two spatial dimensions fractional vortices in charge $2e$ superconductors manifest as pure phase windings of a higher charge superconductor once a non-zero temperature is included; above a critical temperature $T_c$ the superconductivity is lost due to a BKT transition driven by the condensation of these fractional vortices \cite{Mukerjee2006,Leo_09,Berg_2009,Jian2011,Xu2018}. 
 
 We now describe this picture for the candidate superconductors allowed by our statement, starting with the case of $\U(4)$ symmetry.  The Mermin-Wagner theorem forbids the continuous symmetry breaking of the charge $2e$ superconductors, but algebraic order in the $\U(1)$ phase may still persist.  In particular, the unitary charge $2e$ order parameters $\Delta^{\rm S,A}$ that parameterize the manifolds $\U(4)/\Sp(4)$ and $\U(4)/\O(4)$ respectively must be disordered, but the respective submanifolds
 $\U(1)/\Z_4$ and $\U(1)/\Z_8$ retain XY order that melts for
 sufficiently high temperatures through a BKT transition.  The BKT transition is driven by the condensation of half quantum vortices and quarter-quantum vortices for the $\SU(4)$ antisymmetric and symmetric superconductors, respectively.  We may interpet this in terms of higher charge superconductors as follows.  For $T>0$ the charge $2e$ superconductors are unstable: $\Delta^{\rm A,S}$ is disordered.  However, for the antisymmetric case, 
\begin{equation}
    \Delta_{4e} = \varepsilon^{\alpha \beta \gamma \delta} \Delta^A_{\alpha \beta} \Delta^A_{\gamma \delta}
    \label{4efrom2eAsym}
\end{equation}
is an $\SU(4)$ singlet and parameterizes $\U(1)/\Z_4$.  It retains algebraic order and gives rise to a charge $4e$ superconductor for $T>0$.  The $h/4e$ vortices in $\Delta_{4e}$ are the finite temperature remnants of the half quantum vortices in $\Delta^{A}_{2e}$. Similarly the operator 
\begin{equation}
    \Delta_{8e} = \varepsilon^{\alpha \beta \gamma \delta} \varepsilon^{\alpha'\beta'\gamma'\delta'} \Delta^S_{\alpha \alpha'} \Delta^S_{\beta \beta'} \Delta^S_{\gamma \gamma'} \Delta^S_{\delta \delta'}
    \label{4efrom2esym}
\end{equation}
is an $\SU(4)$ singlet built out of $\Delta^S$ that parameterizes $\U(1)/\Z_8$. It can have algebraic order for $T>0$ and host $h/8e$ vortices. 

For intervalley pairing in the $\U(1) \times \SU(2) \times \SU(2)$ symmetric case, the operator
\begin{equation}
    \tilde \Delta_{4e} = \varepsilon^{ab}\varepsilon^{a'b'}  \tilde \Delta_{aa'} \tilde  \Delta_{bb'}
    \label{4efrom2eU2}
\end{equation}
is a singlet with charge $4e$.  Here we used \eqref{DeltaInterValley} to write the order parameter in terms of an arbitrary unitary matrix $\tilde \Delta$.  The unprimed and primed indices transform under the first and second copy of $\SU(2)$, rotations in valley $K$ and $K'$, respectively.  For the antisymmetric representation \eqref{4efrom2eAsym} and \eqref{4efrom2eU2} coincide up to a constant.

It is possible that at $T=0$ these higher charge superconductors are the ground state anyway and there is no charge $2e$ order at $T=0$. The $T>0$ phenomena of higher charge superconductivity are similar in both cases, though the correlation length for charge $2e$ order will diverge exponentially as $T\to 0$ if a charge $2e$ superconductor is the quantum ground state.  
For Josephson junctions with linear extent smaller than the charge $2e$ correlation length the system will behave like a charge $2e$ superconductor.\\

{\bf Symmetry Lowering by Intervalley Hunds Coupling---} Ultimately $\SU(2) \times\ \SU(2)$ symmetry is broken down to its diagonal subgroup of overall spin and charge rotations $\SU(2)$ by intervalley Hunds terms \cite{Zhang2018, TDBGTheory, ScheurerSymm}.  These terms are suppressed by the lattice to moir\'e scale $a/L \ll 1$, and have magnitude $\abs{J_H} \approx 0.2-0.5 \rm{meV}$. Nonzero $J_H$ splits the $\bf{4}$ representation of $\SU(2)\times \SU(2)$ to $\bf{3} \oplus \bf{1}$ of $\SU(2)$, spin triplet and spin singlet respectively \cite{ScheurerSymm}.  The sign of $J_H$ is unknown and determines whether spin triplet or spin singlet states are favored. 

If the singlet state is favored, then at zero temperature there will only be full quantum vortices. The critical temperature for the spin-singlet pairing will be smaller than the critical temperature for charge $4e$ order for sufficiently small $J_H$.  By analogy with the Heisenberg model with a small easy plane bias \cite{Khokhlachev1976,Pelcovits1976,Hikami1980,Cuccoli1995}, we expect the critical temperature to have the following form as $J_H \to 0$
\begin{equation}
    T_{C\text{ singlet,\,} 2e} = \frac{c_1 \rho}{\log \frac{c_2\rho}{J_H}},
    \label{2econdensateJh}
\end{equation}
where $c_{1,2}$ are numbers and $\rho$ is the phase stiffness.  Thus even with small but finite $J_H$  due to the slow variation of the logarithm, there may not be a significant separation of scales between $T_{C\text{ singlet,\,} 2e}$ and   $T_{C\,4e} \propto \rho$.
On the other hand, if the triplet state is favored and the pairing is unitary, half quantum vortices persist.  An extreme non-unitary triplet state is worth mentioning as well, and was considered in Ref. \onlinecite{cornfeld2020spinpolarized}.  Here the order parameter actually parameterizes $\SO(3)$ which has fundamental group $\Z_2$ -- a double vortex is no longer a metastable configuration. Here there is no BKT transition \cite{ScheurerSymm}, however; the $\Z_2$ nature of the vortices implies that there is no composite operator built from the zero temperature order parameter that retains XY order for nonzero temperatures.

Spin orbit coupling at very small scales may also break the overall $\SU(2)$ spin rotation and stabilize a charge $2e$ state with no half quantum vortices at nonzero temperature.  However, spin orbit coupling is extremely weak and so even with a logarithmic critical temperature dependence similar to \eqref{2econdensateJh} we expect that there will be a large window where the charge $2e$ order has melted but the charge $4e$ order is still present with observable half quantum vortices.\\

 \emph{\bf Skyrmions with charge $2e$ and $4e$ and superconductivity}--- Let us now discuss how the different pairing scenarios discussed in this work arise very naturally from the topological mechanism of superconductivity proposed in Ref.~\cite{SkPaper} which we review briefly below. In that work, the authors proposed a topological mechanism for superconductivity based on pairing topological skyrmion textures. They considered a simplified problem where the spin degree of freedom is neglected. This is likely relevant to the extreme non-unitary limit where superconductivity takes place in only one spin species. In this limit, the system consists of two bands in each Chern sector which are labelled by the pseudospin index $\eta$. The low energy states of the system with vanishing Chern number correspond to ferromagnetic order of the pseudospin vector in each Chern sector which are coupled antiferromagnetically between the opposite Chern sectors. A pseudo-spin skyrmion texture in the $C = \pm$ sector carries charge $\pm e$. The antiferromagnetic coupling between the opposite Chern sectors leads to an attractive interaction between a skyrmion in one sector and an antiskyrmion in the opposite sector despite having the same charge. This attraction leads to the formation of a charge $2e$ bound state which leads to superconductivity upon condensation.
 
 Let us now see how this picture changes when we consider the fully spinful model in the absence of flavor symmetry breaking where unitary pairing is expected. In this case, each Chern sector contains four bands labelled by spin and pseudospin. The low energy states with zero total Chern number are specified by filling 2 out of the 4 bands within each sector \cite{KIVCpaper}. Such state is parametrized by  the $4 \times 4$ matrices $Q_\pm$ in the spin-pseudospin space satisfying $Q_\pm^2 = 1$ and $\tr Q_\pm = 0$. Within each Chern sector, $Q_\pm$ parametrizes the manifold $\frac{\U(4)}{\U(2) \times \U(2)}$ which, similar to the 2-sphere $S^2 = \frac{\U(2)}{\U(1) \times \U(1)}$, admits skyrmion textures due to $\pi_2\left(\frac{\U(4)}{\U(2) \times \U(2)} \right) = \Z$. A general skyrmion texture in this manifold intertwines the pseudospin and spin degrees of freedom. We can understand the different types of possible skyrmions by restricting ourselves to some simple limiting cases. One possibility is to consider a pseudospin skyrmion in only one of the spin species. Such a skyrmion can be written as
\beq
Q_+(\br) = \left(\begin{array}{cc}
    \bn_0 \cdot {\boldsymbol \eta} & 0 \\
    0 & \bn_{\rm sk}(\br) \cdot {\boldsymbol \eta}
\end{array} \right)_{\uparrow/\downarrow}
\label{Ske}
\eeq
where $\bn_0$ is some fixed vector whereas $\bn_{\rm sk}(\br)$ describes a skyrmion texture. This skyrmion is the same as the spinless pseudospin skyrmion and will consequently carry an electric charge of $\pm e$. Another more symmetric possibility is to take the same skyrmion in both spin directions so that 
\beq
Q_+(\br) = s_0 \bn_{\rm sk}(\br) \cdot {\boldsymbol \eta}
\label{Sk2e}
\eeq
This skyrmion consists of two copies of the spinless charge $e$ skyrmion and will thus carry a total charge of $2e$. Notice that these Chern skyrmions cannot condense since they feel a net magnus force (effective magnetive field). However, as in the spinless case, the two Chern sectors are antiferromagnetically decoupled by a term of the form $J \tr Q_+ Q_-$ which favors $Q_+ = -Q_-$ leading to an attractive interaction between skyrmions in one Chern sector and antiskyrmions with the same charge in the opposite sector. The resulting bound state feels no net magnetic field and can condense leading to a charge $2e$ or charge $4e$ superconductor. The former is realized by pairing the single spin skyrmions in Eq.~\ref{Ske} which is excepted to break the $\SU(4)$ symmetry while the latter is realized by pairing the spin-singlet skyrmions in Eq.~\ref{Sk2e} which is expected to yield an $\SU(4)$-singlet. A detailed analysis of the symmetry transformation properties of the different skyrmion superconductors will be provided in an upcoming work \cite{SkParton}. 

The discussion above can be substantiated by writing the field theory derived in Ref.~\cite{SkPaper} (although that work focused on the spinless limit, the field theory derived was valid for any number of flavors)
\beq
\L[Q_+, Q_-] = \L_+[Q_+] + \L_-[Q_-] + J \tr Q_+ Q_-
\label{Lpm}
\eeq
with $\L[Q]$ given by
\begin{gather}
   \L_\pm[Q] = \frac{\rho}{2} \tr (\partial_i Q)^2 + i e A_\mu J_{\pm,\mu}[Q] + \L_B[Q] \pm \L_{\rm WZW}[Q], \nonumber \\ J_{\pm,\mu}[Q] = \pm \frac{i \epsilon_{\mu \nu \lambda}}{16 \pi} \tr Q \partial_\nu Q \partial_\lambda Q
\label{LQ0} 
\end{gather}
Here, $\L_B$ is the Berry phase term \cite{MoonMori} whereas $\L_{\rm WZW}$ is a Wess-Zumino-Wittern term which ensures that charge $e$ skyrmions are fermions as expected. This term, which was not included in Ref.~\cite{SkPaper}, will not play a role in our discussion but is generally important to obtain the right symmetry properties as we will show in a subsequent paper \cite{SkParton}. What is important to our current discussion is the skyrmion charge given by integrating $J_{\pm,0}[Q]$. We see that for the single spin skyrmion defined in (\ref{Ske}), the expression in (\ref{LQ0}) simplifies to $\frac{1}{4\pi} \bn_{\rm sk} \cdot (\partial_x \bn_{\rm sk} \times \partial_y \bn_{\rm sk})$ which corresponds to the charge density of a standard spin skyrmion that integrates to 1, thus indicating it is indeed a charge $e$ skyrmion. On the other hand, the spin singlet skyrmion defined in (\ref{Sk2e}) has twice the topological density leading a total electrical charge of $2e$. An effective field theory describing the different insulators and skyrmion superconductors analogous to the $\CP^1$ theory of Ref.~\cite{SkPaper} will be presented in Ref.~\cite{SkParton}.

In conclusion, it is possible to realize the three possibilities -- non-unitary pairing, unitary charge $2e$ and charge $4e$ superconductivity -- within the skyrmion mechanism. The former corresponds to the spinless limit where pairing takes place in a single spin species. The latter two possibilities are realized in the full spinful model by pairing charge $e$ skyrmions (\ref{Ske}) leading to an $\SU(4)$-breaking superconductor or charge $2e$ objects (\ref{Sk2e}) leading to an $\SU(4)$-symmetric superconductor. \\

 \emph{\bf Experimental consequences}--- To discuss the experimental implications of the statement above for TBG, let us first review some experimental facts. In many experiments \cite{PabloSC, PabloNematic, Dean-Young, YoungScreening, EfetovScreening, CaltechSC}, superconductivity is observed in the vicinity of a flavor-symmetry-breaking insulator at $\nu = -2$, which suggests non-unitary pairing. However, there are a few samples \cite{Efetov, EfetovScreening} where superconductivity is observed close to charge neutrality with no signature of flavor symmetry breaking nearby. This is likely to be unitary pairing where cases (ii) or (iii) of our theory should apply. In addition, several groups \cite{EfetovScreening, YoungScreening, CaltechSC} have observed superconductivity in the absence of an insulator close to $\nu = \pm 2$. The absence of insulator is not sufficient to rule out flavor symmetry breaking since it does not exclude the possibility of a flavor-polarized metal. A stronger evidence for flavor polarization  is provided by the presence of a Landau fan emanating for $\nu = \pm2$ indicating a Fermi surface reconstruction. Thus, our general criteria for unitary vs non-unitary pairing can be translated respectively to the absence or presence of a Landau fan emanating from an integer filling in the vicinity of the superconductor. This can be taken as a proxy for flavor symmetry breaking in the parent state. We leave open the unusual scenario where non-unitary pairing develops on top of a flavor symmetric state.  This was observed in $\text{UTe}_{\text{2}}$ \cite{Ran2019} and was argued to occur due to proximity to a ferromagnetic transition \cite{nevidomskyy2020}.
 
Having established the connection to the experimental setup in TBG, let us now investigate the implications of our results. In the vicinity of $\nu = \pm 2$, the most likely flavor polarization is spin which has two possibilities depending on the sign of Hund's coupling. For the ferromagnetic sign of Hund's coupling, the flavor polarized state is a spin ferromagnet which favors a spin-triplet superconductor where pairing takes place in a single spin species. This corresponds to the scenario discussed in Ref.~\cite{cornfeld2020spinpolarized} with only $\Z_2$ vortices. In principle, this implies the superconductor cannot exist at finite temperature and has no BKT transition. This appears hard to reconcile with the robust pairing state observed in experiments. On the other hand, antiferromagnetic Hund's coupling results a non-unitary pairing between anti-aligned spin directions in the two valleys. This state is a superposition of a spin singlet and the $L_z=0$ component of spin-triplet. At finite temperature, $\SU(2)$ rotation symmetry is restored and the spin-triplet component disorders leading to a spin-singlet superconductor. In the absence of spin anisotropies, this appears to be the likely candidate for the experimentally observed superconductivity.

 For superconductors observed in the absence of flavor polarization, we expect unitary pairing with fractional vortices or higher charge superconductivity. Ordinary half quantum vortices are not present in large systems due to logarithmically divergent energies associated with their windings in flavor space.  While the phase winding costs finite energy due to the presence of magnetic flux in the vortex core, there is no physical gauge flux that can cancel the flavor space winding.  However, as discussed above, in two spatial dimensions fractional vortices in charge $2e$
 superconductors manifest as pure phase windings of a higher charge superconductor once a non-zero temperature is included \cite{Mukerjee2006,Jian2011,Xu2018}. 

This has the following implications for the periodicity of the  Fraunhofer patterns in magnetic Josephson junctions or SQUIDS. If the ground state is a charge $2me$ superconductor with $m>1$, this will be observed as a modified periodicity of the Fraunhofer pattern with period $2\pi/m$ for any temperature below the KT transition. On the other hand, if the ground state is a charge $2e$ superconductor, we can only observe the modified periodicity for sufficiently large junctions whose size exceed the charge $2e$ correlation length. The latter decreases with increasing temperature making it likely to observe the modified periodicity at sufficiently large temperatures provided that they are still below the KT temperature.

In mesoscopic annular geometries it may be possible to directly observe half quantum vortices as $h/4e$ jumps in the flux threaded through the hole of the annulus. Mesoscopic geometries are ideal to avoid the logarithmically divergent energy cost associated with the unscreened flavor space winding \cite{Babaev2001, Silaev, Kim2007} and were previously used to observe half quantum vortices \cite{Jang2011,HQVBEC, HQVHe3}.

As discussed above, a Hund's coupling that favors spin singlet pairing will stabilize a unitary charge $2e$ superconductor without half quantum vortices. Because the critical temperature \eqref{2econdensateJh} goes to zero very slowly for small $J_H$, it is likely that experiments done at cold temperatures would not observe such vortices if the Hunds coupling favors singlet pairing. However, the lack of a large separation of scales does not exclude a potentially large temperature window where the singlet Hunds pairing has melted but the charge $4e$ condensate has not.  Within this window one would still observe fractional vortices.

Fractional vortices have previously received much attention due to the promise of Majorana zero modes~\cite{Volovik1999,Read2000,Ivanov2001}. However, these zero modes are specific to $p$-wave pairing where winding in a single spin mimics a full quantum vortex for a spinless $p+ip$ superconductor. Full quantum vortices are not expected to host Majorana zero modes in spinless TBG, and so half quantum vortices in spinful TBG also will not host them. However, both full and half quantum vortices are expected to host multiple protected fermionic zero modes if the pairing is between Chern sectors and valleys. We will discuss these zero modes in more depth in a subsequent paper.\\

\emph{\bf Conclusion}--- In conclusion, we have proven a no-go theorem under mild assumptions which constraints the properties of superconducting condensates in TBG, particularly for regions in the phase diagram where flavor polarization is absent. The main outcome of our analysis is that such regions present ideal hunting grounds for unconventional and exotic types of superconductivity. Our results are independent of the underlying mechanism of superconductivity and are equally applicable to weak coupling BCS-type superconductor as well as strong coupling superconductors. We hope our study will inspire future experiments to look for signatures of fractional vortices and higher charge consendensates in these superconducting regions.\\

\emph{\bf Acknowledgements}--- We thank Igor Babaev and Mathias Scheurer for comments on an early draft of the manuscript. AV was supported by a Simons Investigator award and by the Simons Collaboration on Ultra-Quantum Matter, which is a grant from the Simons Foundation (651440, AV). EK was supported by a Simons Investigator Fellowship, by NSF-DMR 1411343, and by the German National Academy of Sciences Leopoldina through grant LPDS 2018-02 Leopoldina fellowship. 
 
 \bibliography{refs}

\begin{thebibliography}{66}%
\makeatletter
\providecommand \@ifxundefined [1]{%
 \@ifx{#1\undefined}
}%
\providecommand \@ifnum [1]{%
 \ifnum #1\expandafter \@firstoftwo
 \else \expandafter \@secondoftwo
 \fi
}%
\providecommand \@ifx [1]{%
 \ifx #1\expandafter \@firstoftwo
 \else \expandafter \@secondoftwo
 \fi
}%
\providecommand \natexlab [1]{#1}%
\providecommand \enquote  [1]{``#1''}%
\providecommand \bibnamefont  [1]{#1}%
\providecommand \bibfnamefont [1]{#1}%
\providecommand \citenamefont [1]{#1}%
\providecommand \href@noop [0]{\@secondoftwo}%
\providecommand \href [0]{\begingroup \@sanitize@url \@href}%
\providecommand \@href[1]{\@@startlink{#1}\@@href}%
\providecommand \@@href[1]{\endgroup#1\@@endlink}%
\providecommand \@sanitize@url [0]{\catcode `\\12\catcode `\$12\catcode
  `\&12\catcode `\#12\catcode `\^12\catcode `\_12\catcode `\%12\relax}%
\providecommand \@@startlink[1]{}%
\providecommand \@@endlink[0]{}%
\providecommand \url  [0]{\begingroup\@sanitize@url \@url }%
\providecommand \@url [1]{\endgroup\@href {#1}{\urlprefix }}%
\providecommand \urlprefix  [0]{URL }%
\providecommand \Eprint [0]{\href }%
\providecommand \doibase [0]{http://dx.doi.org/}%
\providecommand \selectlanguage [0]{\@gobble}%
\providecommand \bibinfo  [0]{\@secondoftwo}%
\providecommand \bibfield  [0]{\@secondoftwo}%
\providecommand \translation [1]{[#1]}%
\providecommand \BibitemOpen [0]{}%
\providecommand \bibitemStop [0]{}%
\providecommand \bibitemNoStop [0]{.\EOS\space}%
\providecommand \EOS [0]{\spacefactor3000\relax}%
\providecommand \BibitemShut  [1]{\csname bibitem#1\endcsname}%
\let\auto@bib@innerbib\@empty
\bibitem [{\citenamefont {Cao}\ \emph {et~al.}(2018{\natexlab{a}})\citenamefont
  {Cao}, \citenamefont {Fatemi}, \citenamefont {Fang}, \citenamefont
  {Watanabe}, \citenamefont {Taniguchi}, \citenamefont {Kaxiras},\ and\
  \citenamefont {Jarillo-Herrero}}]{PabloSC}%
  \BibitemOpen
  \bibfield  {author} {\bibinfo {author} {\bibfnamefont {Y.}~\bibnamefont
  {Cao}}, \bibinfo {author} {\bibfnamefont {V.}~\bibnamefont {Fatemi}},
  \bibinfo {author} {\bibfnamefont {S.}~\bibnamefont {Fang}}, \bibinfo {author}
  {\bibfnamefont {K.}~\bibnamefont {Watanabe}}, \bibinfo {author}
  {\bibfnamefont {T.}~\bibnamefont {Taniguchi}}, \bibinfo {author}
  {\bibfnamefont {E.}~\bibnamefont {Kaxiras}}, \ and\ \bibinfo {author}
  {\bibfnamefont {P.}~\bibnamefont {Jarillo-Herrero}},\ }\href@noop {}
  {\bibfield  {journal} {\bibinfo  {journal} {Nature}\ }\textbf {\bibinfo
  {volume} {556}},\ \bibinfo {pages} {43} (\bibinfo {year}
  {2018}{\natexlab{a}})}\BibitemShut {NoStop}%
\bibitem [{\citenamefont {Yankowitz}\ \emph {et~al.}(2019)\citenamefont
  {Yankowitz}, \citenamefont {Chen}, \citenamefont {Polshyn}, \citenamefont
  {Zhang}, \citenamefont {Watanabe}, \citenamefont {Taniguchi}, \citenamefont
  {Graf}, \citenamefont {Young},\ and\ \citenamefont {Dean}}]{Dean-Young}%
  \BibitemOpen
  \bibfield  {author} {\bibinfo {author} {\bibfnamefont {M.}~\bibnamefont
  {Yankowitz}}, \bibinfo {author} {\bibfnamefont {S.}~\bibnamefont {Chen}},
  \bibinfo {author} {\bibfnamefont {H.}~\bibnamefont {Polshyn}}, \bibinfo
  {author} {\bibfnamefont {Y.}~\bibnamefont {Zhang}}, \bibinfo {author}
  {\bibfnamefont {K.}~\bibnamefont {Watanabe}}, \bibinfo {author}
  {\bibfnamefont {T.}~\bibnamefont {Taniguchi}}, \bibinfo {author}
  {\bibfnamefont {D.}~\bibnamefont {Graf}}, \bibinfo {author} {\bibfnamefont
  {A.~F.}\ \bibnamefont {Young}}, \ and\ \bibinfo {author} {\bibfnamefont
  {C.~R.}\ \bibnamefont {Dean}},\ }\href@noop {} {\bibfield  {journal}
  {\bibinfo  {journal} {Science}\ ,\ \bibinfo {pages} {1910}} (\bibinfo {year}
  {2019})}\BibitemShut {NoStop}%
\bibitem [{\citenamefont {Lu}\ \emph {et~al.}(2019)\citenamefont {Lu},
  \citenamefont {Stepanov}, \citenamefont {Yang}, \citenamefont {Xie},
  \citenamefont {Aamir}, \citenamefont {Das}, \citenamefont {Urgell},
  \citenamefont {Watanabe}, \citenamefont {Taniguchi}, \citenamefont {Zhang},
  \citenamefont {Bachtold}, \citenamefont {MacDonald},\ and\ \citenamefont
  {Efetov}}]{Efetov}%
  \BibitemOpen
  \bibfield  {author} {\bibinfo {author} {\bibfnamefont {X.}~\bibnamefont
  {Lu}}, \bibinfo {author} {\bibfnamefont {P.}~\bibnamefont {Stepanov}},
  \bibinfo {author} {\bibfnamefont {W.}~\bibnamefont {Yang}}, \bibinfo {author}
  {\bibfnamefont {M.}~\bibnamefont {Xie}}, \bibinfo {author} {\bibfnamefont
  {M.~A.}\ \bibnamefont {Aamir}}, \bibinfo {author} {\bibfnamefont
  {I.}~\bibnamefont {Das}}, \bibinfo {author} {\bibfnamefont {C.}~\bibnamefont
  {Urgell}}, \bibinfo {author} {\bibfnamefont {K.}~\bibnamefont {Watanabe}},
  \bibinfo {author} {\bibfnamefont {T.}~\bibnamefont {Taniguchi}}, \bibinfo
  {author} {\bibfnamefont {G.}~\bibnamefont {Zhang}}, \bibinfo {author}
  {\bibfnamefont {A.}~\bibnamefont {Bachtold}}, \bibinfo {author}
  {\bibfnamefont {A.~H.}\ \bibnamefont {MacDonald}}, \ and\ \bibinfo {author}
  {\bibfnamefont {D.~K.}\ \bibnamefont {Efetov}},\ }\href {\doibase
  10.1038/s41586-019-1695-0} {\bibfield  {journal} {\bibinfo  {journal}
  {Nature}\ }\textbf {\bibinfo {volume} {574}},\ \bibinfo {pages} {653–}
  (\bibinfo {year} {2019})}\BibitemShut {NoStop}%
\bibitem [{\citenamefont {Stepanov}\ \emph {et~al.}(2019)\citenamefont
  {Stepanov}, \citenamefont {Das}, \citenamefont {Lu}, \citenamefont
  {Fahimniya}, \citenamefont {Watanabe}, \citenamefont {Taniguchi},
  \citenamefont {Koppens}, \citenamefont {Lischner}, \citenamefont {Levitov},\
  and\ \citenamefont {Efetov}}]{EfetovScreening}%
  \BibitemOpen
  \bibfield  {author} {\bibinfo {author} {\bibfnamefont {P.}~\bibnamefont
  {Stepanov}}, \bibinfo {author} {\bibfnamefont {I.}~\bibnamefont {Das}},
  \bibinfo {author} {\bibfnamefont {X.}~\bibnamefont {Lu}}, \bibinfo {author}
  {\bibfnamefont {A.}~\bibnamefont {Fahimniya}}, \bibinfo {author}
  {\bibfnamefont {K.}~\bibnamefont {Watanabe}}, \bibinfo {author}
  {\bibfnamefont {T.}~\bibnamefont {Taniguchi}}, \bibinfo {author}
  {\bibfnamefont {F.~H.}\ \bibnamefont {Koppens}}, \bibinfo {author}
  {\bibfnamefont {J.}~\bibnamefont {Lischner}}, \bibinfo {author}
  {\bibfnamefont {L.}~\bibnamefont {Levitov}}, \ and\ \bibinfo {author}
  {\bibfnamefont {D.~K.}\ \bibnamefont {Efetov}},\ }\href@noop {} {\bibfield
  {journal} {\bibinfo  {journal} {arXiv preprint arXiv:1911.09198}\ } (\bibinfo
  {year} {2019})}\BibitemShut {NoStop}%
\bibitem [{\citenamefont {{Saito}}\ \emph {et~al.}(2020)\citenamefont
  {{Saito}}, \citenamefont {{Ge}}, \citenamefont {{Watanabe}}, \citenamefont
  {{Taniguchi}},\ and\ \citenamefont {{Young}}}]{YoungScreening}%
  \BibitemOpen
  \bibfield  {author} {\bibinfo {author} {\bibfnamefont {Y.}~\bibnamefont
  {{Saito}}}, \bibinfo {author} {\bibfnamefont {J.}~\bibnamefont {{Ge}}},
  \bibinfo {author} {\bibfnamefont {K.}~\bibnamefont {{Watanabe}}}, \bibinfo
  {author} {\bibfnamefont {T.}~\bibnamefont {{Taniguchi}}}, \ and\ \bibinfo
  {author} {\bibfnamefont {A.~F.}\ \bibnamefont {{Young}}},\ }\href {\doibase
  10.1038/s41567-020-0928-3} {\bibfield  {journal} {\bibinfo  {journal} {Nature
  Physics}\ }\textbf {\bibinfo {volume} {16}},\ \bibinfo {pages} {926}
  (\bibinfo {year} {2020})}\BibitemShut {NoStop}%
\bibitem [{\citenamefont {Xu}\ and\ \citenamefont {Balents}(2018)}]{Xu2018}%
  \BibitemOpen
  \bibfield  {author} {\bibinfo {author} {\bibfnamefont {C.}~\bibnamefont
  {Xu}}\ and\ \bibinfo {author} {\bibfnamefont {L.}~\bibnamefont {Balents}},\
  }\href {\doibase 10.1103/PhysRevLett.121.087001} {\bibfield  {journal}
  {\bibinfo  {journal} {Phys. Rev. Lett.}\ }\textbf {\bibinfo {volume} {121}},\
  \bibinfo {pages} {087001} (\bibinfo {year} {2018})}\BibitemShut {NoStop}%
\bibitem [{\citenamefont {Lian}\ \emph {et~al.}(2018)\citenamefont {Lian},
  \citenamefont {Wang},\ and\ \citenamefont {Bernevig}}]{phononLianBernevig}%
  \BibitemOpen
  \bibfield  {author} {\bibinfo {author} {\bibfnamefont {B.}~\bibnamefont
  {Lian}}, \bibinfo {author} {\bibfnamefont {Z.}~\bibnamefont {Wang}}, \ and\
  \bibinfo {author} {\bibfnamefont {B.~A.}\ \bibnamefont {Bernevig}},\
  }\href@noop {} {\bibfield  {journal} {\bibinfo  {journal} {arXiv preprint
  arXiv:1807.04382}\ } (\bibinfo {year} {2018})}\BibitemShut {NoStop}%
\bibitem [{\citenamefont {Wu}\ \emph {et~al.}(2018)\citenamefont {Wu},
  \citenamefont {MacDonald},\ and\ \citenamefont {Martin}}]{phononMacDonald}%
  \BibitemOpen
  \bibfield  {author} {\bibinfo {author} {\bibfnamefont {F.}~\bibnamefont
  {Wu}}, \bibinfo {author} {\bibfnamefont {A.~H.}\ \bibnamefont {MacDonald}}, \
  and\ \bibinfo {author} {\bibfnamefont {I.}~\bibnamefont {Martin}},\ }\href
  {\doibase 10.1103/PhysRevLett.121.257001} {\bibfield  {journal} {\bibinfo
  {journal} {Phys. Rev. Lett.}\ }\textbf {\bibinfo {volume} {121}},\ \bibinfo
  {pages} {257001} (\bibinfo {year} {2018})}\BibitemShut {NoStop}%
\bibitem [{\citenamefont {You}\ and\ \citenamefont
  {Vishwanath}(2019)}]{YouVishwanath}%
  \BibitemOpen
  \bibfield  {author} {\bibinfo {author} {\bibfnamefont {Y.-Z.}\ \bibnamefont
  {You}}\ and\ \bibinfo {author} {\bibfnamefont {A.}~\bibnamefont
  {Vishwanath}},\ }\href {\doibase 10.1038/s41535-019-0153-4} {\bibfield
  {journal} {\bibinfo  {journal} {npj Quantum Materials}\ }\textbf {\bibinfo
  {volume} {4}},\ \bibinfo {pages} {16} (\bibinfo {year} {2019})}\BibitemShut
  {NoStop}%
\bibitem [{\citenamefont {Isobe}\ \emph {et~al.}(2018)\citenamefont {Isobe},
  \citenamefont {Yuan},\ and\ \citenamefont {Fu}}]{IsobeFuSC}%
  \BibitemOpen
  \bibfield  {author} {\bibinfo {author} {\bibfnamefont {H.}~\bibnamefont
  {Isobe}}, \bibinfo {author} {\bibfnamefont {N.~F.~Q.}\ \bibnamefont {Yuan}},
  \ and\ \bibinfo {author} {\bibfnamefont {L.}~\bibnamefont {Fu}},\ }\href
  {\doibase 10.1103/PhysRevX.8.041041} {\bibfield  {journal} {\bibinfo
  {journal} {Phys. Rev. X}\ }\textbf {\bibinfo {volume} {8}},\ \bibinfo {pages}
  {041041} (\bibinfo {year} {2018})}\BibitemShut {NoStop}%
\bibitem [{\citenamefont {Classen}\ \emph {et~al.}(2020)\citenamefont
  {Classen}, \citenamefont {Chubukov}, \citenamefont {Honerkamp},\ and\
  \citenamefont {Scherer}}]{ChubukovVanHove}%
  \BibitemOpen
  \bibfield  {author} {\bibinfo {author} {\bibfnamefont {L.}~\bibnamefont
  {Classen}}, \bibinfo {author} {\bibfnamefont {A.~V.}\ \bibnamefont
  {Chubukov}}, \bibinfo {author} {\bibfnamefont {C.}~\bibnamefont {Honerkamp}},
  \ and\ \bibinfo {author} {\bibfnamefont {M.~M.}\ \bibnamefont {Scherer}},\
  }\href {\doibase 10.1103/PhysRevB.102.125141} {\bibfield  {journal} {\bibinfo
   {journal} {Phys. Rev. B}\ }\textbf {\bibinfo {volume} {102}},\ \bibinfo
  {pages} {125141} (\bibinfo {year} {2020})}\BibitemShut {NoStop}%
\bibitem [{\citenamefont {Chichinadze}\ \emph {et~al.}(2020)\citenamefont
  {Chichinadze}, \citenamefont {Classen},\ and\ \citenamefont
  {Chubukov}}]{ChubukovTBG}%
  \BibitemOpen
  \bibfield  {author} {\bibinfo {author} {\bibfnamefont {D.~V.}\ \bibnamefont
  {Chichinadze}}, \bibinfo {author} {\bibfnamefont {L.}~\bibnamefont
  {Classen}}, \ and\ \bibinfo {author} {\bibfnamefont {A.~V.}\ \bibnamefont
  {Chubukov}},\ }\href {\doibase 10.1103/PhysRevB.102.125120} {\bibfield
  {journal} {\bibinfo  {journal} {Phys. Rev. B}\ }\textbf {\bibinfo {volume}
  {102}},\ \bibinfo {pages} {125120} (\bibinfo {year} {2020})}\BibitemShut
  {NoStop}%
\bibitem [{\citenamefont {Fernandes}\ and\ \citenamefont
  {Venderbos}(2020)}]{FernandesVanderbos}%
  \BibitemOpen
  \bibfield  {author} {\bibinfo {author} {\bibfnamefont {R.~M.}\ \bibnamefont
  {Fernandes}}\ and\ \bibinfo {author} {\bibfnamefont {J.~W.~F.}\ \bibnamefont
  {Venderbos}},\ }\href {\doibase 10.1126/sciadv.aba8834} {\bibfield  {journal}
  {\bibinfo  {journal} {Science Advances}\ }\textbf {\bibinfo {volume} {6}}
  (\bibinfo {year} {2020}),\ 10.1126/sciadv.aba8834},\ \Eprint
  {http://arxiv.org/abs/https://advances.sciencemag.org/content/6/32/eaba8834.full.pdf}
  {https://advances.sciencemag.org/content/6/32/eaba8834.full.pdf} \BibitemShut
  {NoStop}%
\bibitem [{\citenamefont {{Wang}}\ \emph {et~al.}(2020)\citenamefont {{Wang}},
  \citenamefont {{Kang}},\ and\ \citenamefont {{Fernandes}}}]{KangFernandes}%
  \BibitemOpen
  \bibfield  {author} {\bibinfo {author} {\bibfnamefont {Y.}~\bibnamefont
  {{Wang}}}, \bibinfo {author} {\bibfnamefont {J.}~\bibnamefont {{Kang}}}, \
  and\ \bibinfo {author} {\bibfnamefont {R.~M.}\ \bibnamefont {{Fernandes}}},\
  }\href@noop {} {\bibfield  {journal} {\bibinfo  {journal} {arXiv e-prints}\
  ,\ \bibinfo {pages} {2009.01237}} (\bibinfo {year} {2020})}\BibitemShut
  {NoStop}%
\bibitem [{\citenamefont {{Kozii}}\ \emph {et~al.}(2020)\citenamefont
  {{Kozii}}, \citenamefont {{Zaletel}},\ and\ \citenamefont
  {{Bultinck}}}]{Kozii}%
  \BibitemOpen
  \bibfield  {author} {\bibinfo {author} {\bibfnamefont {V.}~\bibnamefont
  {{Kozii}}}, \bibinfo {author} {\bibfnamefont {M.~P.}\ \bibnamefont
  {{Zaletel}}}, \ and\ \bibinfo {author} {\bibfnamefont {N.}~\bibnamefont
  {{Bultinck}}},\ }\href@noop {} {\bibfield  {journal} {\bibinfo  {journal}
  {arXiv e-prints}\ ,\ \bibinfo {pages} {2005.12961}} (\bibinfo {year}
  {2020})}\BibitemShut {NoStop}%
\bibitem [{\citenamefont {Khalaf}\ \emph {et~al.}(2020)\citenamefont {Khalaf},
  \citenamefont {Chatterjee}, \citenamefont {Bultinck}, \citenamefont
  {Zaletel},\ and\ \citenamefont {Vishwanath}}]{SkPaper}%
  \BibitemOpen
  \bibfield  {author} {\bibinfo {author} {\bibfnamefont {E.}~\bibnamefont
  {Khalaf}}, \bibinfo {author} {\bibfnamefont {S.}~\bibnamefont {Chatterjee}},
  \bibinfo {author} {\bibfnamefont {N.}~\bibnamefont {Bultinck}}, \bibinfo
  {author} {\bibfnamefont {M.~P.}\ \bibnamefont {Zaletel}}, \ and\ \bibinfo
  {author} {\bibfnamefont {A.}~\bibnamefont {Vishwanath}},\ }\href@noop {}
  {\bibfield  {journal} {\bibinfo  {journal} {arXiv preprint arXiv:2004.00638}\
  } (\bibinfo {year} {2020})}\BibitemShut {NoStop}%
\bibitem [{\citenamefont {Chatterjee}\ \emph {et~al.}(2020)\citenamefont
  {Chatterjee}, \citenamefont {Ippoliti},\ and\ \citenamefont
  {Zaletel}}]{SkDMRG}%
  \BibitemOpen
  \bibfield  {author} {\bibinfo {author} {\bibfnamefont {S.}~\bibnamefont
  {Chatterjee}}, \bibinfo {author} {\bibfnamefont {M.}~\bibnamefont
  {Ippoliti}}, \ and\ \bibinfo {author} {\bibfnamefont {M.~P.}\ \bibnamefont
  {Zaletel}},\ }\href@noop {} {\bibfield  {journal} {\bibinfo  {journal} {arXiv
  preprint arXiv:2010.01144}\ } (\bibinfo {year} {2020})}\BibitemShut {NoStop}%
\bibitem [{\citenamefont {Christos}\ \emph {et~al.}(2020)\citenamefont
  {Christos}, \citenamefont {Sachdev},\ and\ \citenamefont
  {Scheurer}}]{Christos}%
  \BibitemOpen
  \bibfield  {author} {\bibinfo {author} {\bibfnamefont {M.}~\bibnamefont
  {Christos}}, \bibinfo {author} {\bibfnamefont {S.}~\bibnamefont {Sachdev}}, \
  and\ \bibinfo {author} {\bibfnamefont {M.~S.}\ \bibnamefont {Scheurer}},\
  }\href {\doibase 10.1073/pnas.2014691117} {\bibfield  {journal} {\bibinfo
  {journal} {Proceedings of the National Academy of Sciences}\ }\textbf
  {\bibinfo {volume} {117}},\ \bibinfo {pages} {29543} (\bibinfo {year}
  {2020})},\ \Eprint
  {http://arxiv.org/abs/https://www.pnas.org/content/117/47/29543.full.pdf}
  {https://www.pnas.org/content/117/47/29543.full.pdf} \BibitemShut {NoStop}%
\bibitem [{\citenamefont {Lian}\ \emph {et~al.}(2020)\citenamefont {Lian},
  \citenamefont {Xie},\ and\ \citenamefont {Bernevig}}]{BernevigTBGTopology}%
  \BibitemOpen
  \bibfield  {author} {\bibinfo {author} {\bibfnamefont {B.}~\bibnamefont
  {Lian}}, \bibinfo {author} {\bibfnamefont {F.}~\bibnamefont {Xie}}, \ and\
  \bibinfo {author} {\bibfnamefont {B.~A.}\ \bibnamefont {Bernevig}},\ }\href
  {\doibase 10.1103/PhysRevB.102.041402} {\bibfield  {journal} {\bibinfo
  {journal} {Phys. Rev. B}\ }\textbf {\bibinfo {volume} {102}},\ \bibinfo
  {pages} {041402} (\bibinfo {year} {2020})}\BibitemShut {NoStop}%
\bibitem [{\citenamefont {Bernevig}\ \emph {et~al.}(2020)\citenamefont
  {Bernevig}, \citenamefont {Lian}, \citenamefont {Cowsik}, \citenamefont
  {Xie}, \citenamefont {Regnault},\ and\ \citenamefont {Song}}]{TBGV}%
  \BibitemOpen
  \bibfield  {author} {\bibinfo {author} {\bibfnamefont {B.~A.}\ \bibnamefont
  {Bernevig}}, \bibinfo {author} {\bibfnamefont {B.}~\bibnamefont {Lian}},
  \bibinfo {author} {\bibfnamefont {A.}~\bibnamefont {Cowsik}}, \bibinfo
  {author} {\bibfnamefont {F.}~\bibnamefont {Xie}}, \bibinfo {author}
  {\bibfnamefont {N.}~\bibnamefont {Regnault}}, \ and\ \bibinfo {author}
  {\bibfnamefont {Z.-D.}\ \bibnamefont {Song}},\ }\href@noop {} {\bibfield
  {journal} {\bibinfo  {journal} {arXiv preprint arXiv:2009.14200}\ } (\bibinfo
  {year} {2020})}\BibitemShut {NoStop}%
\bibitem [{\citenamefont {Scheurer}\ and\ \citenamefont
  {Samajdar}(2020)}]{ScheurerSymm}%
  \BibitemOpen
  \bibfield  {author} {\bibinfo {author} {\bibfnamefont {M.~S.}\ \bibnamefont
  {Scheurer}}\ and\ \bibinfo {author} {\bibfnamefont {R.}~\bibnamefont
  {Samajdar}},\ }\href {\doibase 10.1103/PhysRevResearch.2.033062} {\bibfield
  {journal} {\bibinfo  {journal} {Phys. Rev. Research}\ }\textbf {\bibinfo
  {volume} {2}},\ \bibinfo {pages} {033062} (\bibinfo {year}
  {2020})}\BibitemShut {NoStop}%
\bibitem [{\citenamefont {Leggett}(1975)}]{Leggett75}%
  \BibitemOpen
  \bibfield  {author} {\bibinfo {author} {\bibfnamefont {A.~J.}\ \bibnamefont
  {Leggett}},\ }\href {\doibase 10.1103/RevModPhys.47.331} {\bibfield
  {journal} {\bibinfo  {journal} {Rev. Mod. Phys.}\ }\textbf {\bibinfo {volume}
  {47}},\ \bibinfo {pages} {331} (\bibinfo {year} {1975})}\BibitemShut
  {NoStop}%
\bibitem [{\citenamefont {Salomaa}\ and\ \citenamefont
  {Volovik}(1987)}]{Volovik87}%
  \BibitemOpen
  \bibfield  {author} {\bibinfo {author} {\bibfnamefont {M.~M.}\ \bibnamefont
  {Salomaa}}\ and\ \bibinfo {author} {\bibfnamefont {G.~E.}\ \bibnamefont
  {Volovik}},\ }\href {\doibase 10.1103/RevModPhys.59.533} {\bibfield
  {journal} {\bibinfo  {journal} {Rev. Mod. Phys.}\ }\textbf {\bibinfo {volume}
  {59}},\ \bibinfo {pages} {533} (\bibinfo {year} {1987})}\BibitemShut
  {NoStop}%
\bibitem [{\citenamefont {Bultinck}\ \emph {et~al.}(2020)\citenamefont
  {Bultinck}, \citenamefont {Khalaf}, \citenamefont {Liu}, \citenamefont
  {Chatterjee}, \citenamefont {Vishwanath},\ and\ \citenamefont
  {Zaletel}}]{KIVCpaper}%
  \BibitemOpen
  \bibfield  {author} {\bibinfo {author} {\bibfnamefont {N.}~\bibnamefont
  {Bultinck}}, \bibinfo {author} {\bibfnamefont {E.}~\bibnamefont {Khalaf}},
  \bibinfo {author} {\bibfnamefont {S.}~\bibnamefont {Liu}}, \bibinfo {author}
  {\bibfnamefont {S.}~\bibnamefont {Chatterjee}}, \bibinfo {author}
  {\bibfnamefont {A.}~\bibnamefont {Vishwanath}}, \ and\ \bibinfo {author}
  {\bibfnamefont {M.~P.}\ \bibnamefont {Zaletel}},\ }\href {\doibase
  10.1103/PhysRevX.10.031034} {\bibfield  {journal} {\bibinfo  {journal} {Phys.
  Rev. X}\ }\textbf {\bibinfo {volume} {10}},\ \bibinfo {pages} {031034}
  (\bibinfo {year} {2020})}\BibitemShut {NoStop}%
\bibitem [{\citenamefont {Kang}\ and\ \citenamefont
  {Vafek}(2019)}]{KangVafekPRL}%
  \BibitemOpen
  \bibfield  {author} {\bibinfo {author} {\bibfnamefont {J.}~\bibnamefont
  {Kang}}\ and\ \bibinfo {author} {\bibfnamefont {O.}~\bibnamefont {Vafek}},\
  }\href {\doibase 10.1103/PhysRevLett.122.246401} {\bibfield  {journal}
  {\bibinfo  {journal} {Phys. Rev. Lett.}\ }\textbf {\bibinfo {volume} {122}},\
  \bibinfo {pages} {246401} (\bibinfo {year} {2019})}\BibitemShut {NoStop}%
\bibitem [{\citenamefont {Seo}\ \emph {et~al.}(2019)\citenamefont {Seo},
  \citenamefont {Kotov},\ and\ \citenamefont {Uchoa}}]{Uchoa}%
  \BibitemOpen
  \bibfield  {author} {\bibinfo {author} {\bibfnamefont {K.}~\bibnamefont
  {Seo}}, \bibinfo {author} {\bibfnamefont {V.~N.}\ \bibnamefont {Kotov}}, \
  and\ \bibinfo {author} {\bibfnamefont {B.}~\bibnamefont {Uchoa}},\ }\href
  {\doibase 10.1103/PhysRevLett.122.246402} {\bibfield  {journal} {\bibinfo
  {journal} {Phys. Rev. Lett.}\ }\textbf {\bibinfo {volume} {122}},\ \bibinfo
  {pages} {246402} (\bibinfo {year} {2019})}\BibitemShut {NoStop}%
\bibitem [{\citenamefont {Vafek}\ and\ \citenamefont
  {Kang}(2020)}]{VafekKangRG}%
  \BibitemOpen
  \bibfield  {author} {\bibinfo {author} {\bibfnamefont {O.}~\bibnamefont
  {Vafek}}\ and\ \bibinfo {author} {\bibfnamefont {J.}~\bibnamefont {Kang}},\
  }\href@noop {} {\bibfield  {journal} {\bibinfo  {journal} {arXiv preprint
  arXiv:2009.09413}\ } (\bibinfo {year} {2020})}\BibitemShut {NoStop}%
\bibitem [{\citenamefont {Tarnopolsky}\ \emph {et~al.}(2019)\citenamefont
  {Tarnopolsky}, \citenamefont {Kruchkov},\ and\ \citenamefont
  {Vishwanath}}]{Tarnopolsky}%
  \BibitemOpen
  \bibfield  {author} {\bibinfo {author} {\bibfnamefont {G.}~\bibnamefont
  {Tarnopolsky}}, \bibinfo {author} {\bibfnamefont {A.~J.}\ \bibnamefont
  {Kruchkov}}, \ and\ \bibinfo {author} {\bibfnamefont {A.}~\bibnamefont
  {Vishwanath}},\ }\href {\doibase 10.1103/PhysRevLett.122.106405} {\bibfield
  {journal} {\bibinfo  {journal} {Phys. Rev. Lett.}\ }\textbf {\bibinfo
  {volume} {122}},\ \bibinfo {pages} {106405} (\bibinfo {year}
  {2019})}\BibitemShut {NoStop}%
\bibitem [{\citenamefont {Huy}\ \emph {et~al.}(2007)\citenamefont {Huy},
  \citenamefont {Gasparini}, \citenamefont {de~Nijs}, \citenamefont {Huang},
  \citenamefont {Klaasse}, \citenamefont {Gortenmulder}, \citenamefont
  {de~Visser}, \citenamefont {Hamann}, \citenamefont {G\"orlach},\ and\
  \citenamefont {L\"ohneysen}}]{UCoGe}%
  \BibitemOpen
  \bibfield  {author} {\bibinfo {author} {\bibfnamefont {N.~T.}\ \bibnamefont
  {Huy}}, \bibinfo {author} {\bibfnamefont {A.}~\bibnamefont {Gasparini}},
  \bibinfo {author} {\bibfnamefont {D.~E.}\ \bibnamefont {de~Nijs}}, \bibinfo
  {author} {\bibfnamefont {Y.}~\bibnamefont {Huang}}, \bibinfo {author}
  {\bibfnamefont {J.~C.~P.}\ \bibnamefont {Klaasse}}, \bibinfo {author}
  {\bibfnamefont {T.}~\bibnamefont {Gortenmulder}}, \bibinfo {author}
  {\bibfnamefont {A.}~\bibnamefont {de~Visser}}, \bibinfo {author}
  {\bibfnamefont {A.}~\bibnamefont {Hamann}}, \bibinfo {author} {\bibfnamefont
  {T.}~\bibnamefont {G\"orlach}}, \ and\ \bibinfo {author} {\bibfnamefont
  {H.~v.}\ \bibnamefont {L\"ohneysen}},\ }\href {\doibase
  10.1103/PhysRevLett.99.067006} {\bibfield  {journal} {\bibinfo  {journal}
  {Phys. Rev. Lett.}\ }\textbf {\bibinfo {volume} {99}},\ \bibinfo {pages}
  {067006} (\bibinfo {year} {2007})}\BibitemShut {NoStop}%
\bibitem [{\citenamefont {Aoki}\ and\ \citenamefont
  {Flouquet}(2012)}]{Uranium}%
  \BibitemOpen
  \bibfield  {author} {\bibinfo {author} {\bibfnamefont {D.}~\bibnamefont
  {Aoki}}\ and\ \bibinfo {author} {\bibfnamefont {J.}~\bibnamefont
  {Flouquet}},\ }\href {\doibase 10.1143/JPSJ.81.011003} {\bibfield  {journal}
  {\bibinfo  {journal} {Journal of the Physical Society of Japan}\ }\textbf
  {\bibinfo {volume} {81}},\ \bibinfo {pages} {011003} (\bibinfo {year}
  {2012})},\ \Eprint
  {http://arxiv.org/abs/https://doi.org/10.1143/JPSJ.81.011003}
  {https://doi.org/10.1143/JPSJ.81.011003} \BibitemShut {NoStop}%
\bibitem [{\citenamefont {Nevidomskyy}(2020{\natexlab{a}})}]{Nevidomskyy}%
  \BibitemOpen
  \bibfield  {author} {\bibinfo {author} {\bibfnamefont {A.~H.}\ \bibnamefont
  {Nevidomskyy}},\ }\href@noop {} {\bibfield  {journal} {\bibinfo  {journal}
  {arXiv preprint arXiv:2001.02699}\ } (\bibinfo {year}
  {2020}{\natexlab{a}})}\BibitemShut {NoStop}%
\bibitem [{\citenamefont {Ran}\ \emph {et~al.}(2019{\natexlab{a}})\citenamefont
  {Ran}, \citenamefont {Eckberg}, \citenamefont {Ding}, \citenamefont
  {Furukawa}, \citenamefont {Metz}, \citenamefont {Saha}, \citenamefont {Liu},
  \citenamefont {Zic}, \citenamefont {Kim}, \citenamefont {Paglione},\ and\
  \citenamefont {Butch}}]{UTeScience}%
  \BibitemOpen
  \bibfield  {author} {\bibinfo {author} {\bibfnamefont {S.}~\bibnamefont
  {Ran}}, \bibinfo {author} {\bibfnamefont {C.}~\bibnamefont {Eckberg}},
  \bibinfo {author} {\bibfnamefont {Q.-P.}\ \bibnamefont {Ding}}, \bibinfo
  {author} {\bibfnamefont {Y.}~\bibnamefont {Furukawa}}, \bibinfo {author}
  {\bibfnamefont {T.}~\bibnamefont {Metz}}, \bibinfo {author} {\bibfnamefont
  {S.~R.}\ \bibnamefont {Saha}}, \bibinfo {author} {\bibfnamefont {I.-L.}\
  \bibnamefont {Liu}}, \bibinfo {author} {\bibfnamefont {M.}~\bibnamefont
  {Zic}}, \bibinfo {author} {\bibfnamefont {H.}~\bibnamefont {Kim}}, \bibinfo
  {author} {\bibfnamefont {J.}~\bibnamefont {Paglione}}, \ and\ \bibinfo
  {author} {\bibfnamefont {N.~P.}\ \bibnamefont {Butch}},\ }\href {\doibase
  10.1126/science.aav8645} {\bibfield  {journal} {\bibinfo  {journal}
  {Science}\ }\textbf {\bibinfo {volume} {365}},\ \bibinfo {pages} {684}
  (\bibinfo {year} {2019}{\natexlab{a}})},\ \Eprint
  {http://arxiv.org/abs/https://science.sciencemag.org/content/365/6454/684.full.pdf}
  {https://science.sciencemag.org/content/365/6454/684.full.pdf} \BibitemShut
  {NoStop}%
\bibitem [{\citenamefont {Autti}\ \emph {et~al.}(2016)\citenamefont {Autti},
  \citenamefont {Dmitriev}, \citenamefont {M\"akinen}, \citenamefont
  {Soldatov}, \citenamefont {Volovik}, \citenamefont {Yudin}, \citenamefont
  {Zavjalov},\ and\ \citenamefont {Eltsov}}]{HQVHe3}%
  \BibitemOpen
  \bibfield  {author} {\bibinfo {author} {\bibfnamefont {S.}~\bibnamefont
  {Autti}}, \bibinfo {author} {\bibfnamefont {V.~V.}\ \bibnamefont {Dmitriev}},
  \bibinfo {author} {\bibfnamefont {J.~T.}\ \bibnamefont {M\"akinen}}, \bibinfo
  {author} {\bibfnamefont {A.~A.}\ \bibnamefont {Soldatov}}, \bibinfo {author}
  {\bibfnamefont {G.~E.}\ \bibnamefont {Volovik}}, \bibinfo {author}
  {\bibfnamefont {A.~N.}\ \bibnamefont {Yudin}}, \bibinfo {author}
  {\bibfnamefont {V.~V.}\ \bibnamefont {Zavjalov}}, \ and\ \bibinfo {author}
  {\bibfnamefont {V.~B.}\ \bibnamefont {Eltsov}},\ }\href {\doibase
  10.1103/PhysRevLett.117.255301} {\bibfield  {journal} {\bibinfo  {journal}
  {Phys. Rev. Lett.}\ }\textbf {\bibinfo {volume} {117}},\ \bibinfo {pages}
  {255301} (\bibinfo {year} {2016})}\BibitemShut {NoStop}%
\bibitem [{\citenamefont {Mackenzie}\ and\ \citenamefont
  {Maeno}(2003)}]{SrRuO}%
  \BibitemOpen
  \bibfield  {author} {\bibinfo {author} {\bibfnamefont {A.~P.}\ \bibnamefont
  {Mackenzie}}\ and\ \bibinfo {author} {\bibfnamefont {Y.}~\bibnamefont
  {Maeno}},\ }\href {\doibase 10.1103/RevModPhys.75.657} {\bibfield  {journal}
  {\bibinfo  {journal} {Rev. Mod. Phys.}\ }\textbf {\bibinfo {volume} {75}},\
  \bibinfo {pages} {657} (\bibinfo {year} {2003})}\BibitemShut {NoStop}%
\bibitem [{\citenamefont {Babaev}(2002)}]{Babaev2001}%
  \BibitemOpen
  \bibfield  {author} {\bibinfo {author} {\bibfnamefont {E.}~\bibnamefont
  {Babaev}},\ }\href {\doibase 10.1103/PhysRevLett.89.067001} {\bibfield
  {journal} {\bibinfo  {journal} {Phys. Rev. Lett.}\ }\textbf {\bibinfo
  {volume} {89}},\ \bibinfo {pages} {067001} (\bibinfo {year}
  {2002})}\BibitemShut {NoStop}%
\bibitem [{\citenamefont {Silaev}(2011)}]{Silaev}%
  \BibitemOpen
  \bibfield  {author} {\bibinfo {author} {\bibfnamefont {M.~A.}\ \bibnamefont
  {Silaev}},\ }\href {\doibase 10.1103/PhysRevB.83.144519} {\bibfield
  {journal} {\bibinfo  {journal} {Phys. Rev. B}\ }\textbf {\bibinfo {volume}
  {83}},\ \bibinfo {pages} {144519} (\bibinfo {year} {2011})}\BibitemShut
  {NoStop}%
\bibitem [{\citenamefont {Chung}\ \emph {et~al.}(2007)\citenamefont {Chung},
  \citenamefont {Bluhm},\ and\ \citenamefont {Kim}}]{Kim2007}%
  \BibitemOpen
  \bibfield  {author} {\bibinfo {author} {\bibfnamefont {S.~B.}\ \bibnamefont
  {Chung}}, \bibinfo {author} {\bibfnamefont {H.}~\bibnamefont {Bluhm}}, \ and\
  \bibinfo {author} {\bibfnamefont {E.-A.}\ \bibnamefont {Kim}},\ }\href
  {\doibase 10.1103/PhysRevLett.99.197002} {\bibfield  {journal} {\bibinfo
  {journal} {Phys. Rev. Lett.}\ }\textbf {\bibinfo {volume} {99}},\ \bibinfo
  {pages} {197002} (\bibinfo {year} {2007})}\BibitemShut {NoStop}%
\bibitem [{\citenamefont {Liu}\ \emph {et~al.}(2019)\citenamefont {Liu},
  \citenamefont {Khalaf}, \citenamefont {Lee},\ and\ \citenamefont
  {Vishwanath}}]{ShangHF}%
  \BibitemOpen
  \bibfield  {author} {\bibinfo {author} {\bibfnamefont {S.}~\bibnamefont
  {Liu}}, \bibinfo {author} {\bibfnamefont {E.}~\bibnamefont {Khalaf}},
  \bibinfo {author} {\bibfnamefont {J.~Y.}\ \bibnamefont {Lee}}, \ and\
  \bibinfo {author} {\bibfnamefont {A.}~\bibnamefont {Vishwanath}},\
  }\href@noop {} {\bibfield  {journal} {\bibinfo  {journal} {arXiv preprint
  arXiv:1905.07409}\ } (\bibinfo {year} {2019})}\BibitemShut {NoStop}%
\bibitem [{\citenamefont {Xie}\ and\ \citenamefont
  {MacDonald}(2020)}]{MacdonaldHF}%
  \BibitemOpen
  \bibfield  {author} {\bibinfo {author} {\bibfnamefont {M.}~\bibnamefont
  {Xie}}\ and\ \bibinfo {author} {\bibfnamefont {A.~H.}\ \bibnamefont
  {MacDonald}},\ }\href {\doibase 10.1103/PhysRevLett.124.097601} {\bibfield
  {journal} {\bibinfo  {journal} {Phys. Rev. Lett.}\ }\textbf {\bibinfo
  {volume} {124}},\ \bibinfo {pages} {097601} (\bibinfo {year}
  {2020})}\BibitemShut {NoStop}%
\bibitem [{\citenamefont {Cao}\ \emph {et~al.}(2018{\natexlab{b}})\citenamefont
  {Cao}, \citenamefont {Fatemi}, \citenamefont {Demir}, \citenamefont {Fang},
  \citenamefont {Tomarken}, \citenamefont {Luo}, \citenamefont
  {Sanchez-Yamagishi}, \citenamefont {Watanabe}, \citenamefont {Taniguchi},
  \citenamefont {Kaxiras} \emph {et~al.}}]{PabloMott}%
  \BibitemOpen
  \bibfield  {author} {\bibinfo {author} {\bibfnamefont {Y.}~\bibnamefont
  {Cao}}, \bibinfo {author} {\bibfnamefont {V.}~\bibnamefont {Fatemi}},
  \bibinfo {author} {\bibfnamefont {A.}~\bibnamefont {Demir}}, \bibinfo
  {author} {\bibfnamefont {S.}~\bibnamefont {Fang}}, \bibinfo {author}
  {\bibfnamefont {S.~L.}\ \bibnamefont {Tomarken}}, \bibinfo {author}
  {\bibfnamefont {J.~Y.}\ \bibnamefont {Luo}}, \bibinfo {author} {\bibfnamefont
  {J.~D.}\ \bibnamefont {Sanchez-Yamagishi}}, \bibinfo {author} {\bibfnamefont
  {K.}~\bibnamefont {Watanabe}}, \bibinfo {author} {\bibfnamefont
  {T.}~\bibnamefont {Taniguchi}}, \bibinfo {author} {\bibfnamefont
  {E.}~\bibnamefont {Kaxiras}},  \emph {et~al.},\ }\href@noop {} {\bibfield
  {journal} {\bibinfo  {journal} {Nature}\ }\textbf {\bibinfo {volume} {556}},\
  \bibinfo {pages} {80} (\bibinfo {year} {2018}{\natexlab{b}})}\BibitemShut
  {NoStop}%
\bibitem [{\citenamefont {Arora}\ \emph {et~al.}(2020)\citenamefont {Arora},
  \citenamefont {Polski}, \citenamefont {Zhang}, \citenamefont {Thomson},
  \citenamefont {Choi}, \citenamefont {Kim}, \citenamefont {Lin}, \citenamefont
  {Wilson}, \citenamefont {Xu}, \citenamefont {Chu}, \citenamefont {Watanabe},
  \citenamefont {Taniguchi}, \citenamefont {Alicea},\ and\ \citenamefont
  {Nadj-Perge}}]{CaltechSC}%
  \BibitemOpen
  \bibfield  {author} {\bibinfo {author} {\bibfnamefont {H.~S.}\ \bibnamefont
  {Arora}}, \bibinfo {author} {\bibfnamefont {R.}~\bibnamefont {Polski}},
  \bibinfo {author} {\bibfnamefont {Y.}~\bibnamefont {Zhang}}, \bibinfo
  {author} {\bibfnamefont {A.}~\bibnamefont {Thomson}}, \bibinfo {author}
  {\bibfnamefont {Y.}~\bibnamefont {Choi}}, \bibinfo {author} {\bibfnamefont
  {H.}~\bibnamefont {Kim}}, \bibinfo {author} {\bibfnamefont {Z.}~\bibnamefont
  {Lin}}, \bibinfo {author} {\bibfnamefont {I.~Z.}\ \bibnamefont {Wilson}},
  \bibinfo {author} {\bibfnamefont {X.}~\bibnamefont {Xu}}, \bibinfo {author}
  {\bibfnamefont {J.-H.}\ \bibnamefont {Chu}}, \bibinfo {author} {\bibfnamefont
  {K.}~\bibnamefont {Watanabe}}, \bibinfo {author} {\bibfnamefont
  {T.}~\bibnamefont {Taniguchi}}, \bibinfo {author} {\bibfnamefont
  {J.}~\bibnamefont {Alicea}}, \ and\ \bibinfo {author} {\bibfnamefont
  {S.}~\bibnamefont {Nadj-Perge}},\ }\href@noop {} {\emph {\bibinfo {title}
  {{Superconductivity without insulating states in twisted bi-layer graphene
  stabilized by monolayer WSe 2}}}},\ \bibinfo {type} {Tech. Rep.}\ (\bibinfo
  {year} {2020})\ \Eprint {http://arxiv.org/abs/2002.03003v1}
  {arXiv:2002.03003v1} \BibitemShut {NoStop}%
\bibitem [{\citenamefont {Park}\ \emph {et~al.}(2020)\citenamefont {Park},
  \citenamefont {Cao}, \citenamefont {Watanabe}, \citenamefont {Taniguchi},\
  and\ \citenamefont {Jarillo-Herrero}}]{PabloTrilayer}%
  \BibitemOpen
  \bibfield  {author} {\bibinfo {author} {\bibfnamefont {J.~M.}\ \bibnamefont
  {Park}}, \bibinfo {author} {\bibfnamefont {Y.}~\bibnamefont {Cao}}, \bibinfo
  {author} {\bibfnamefont {K.}~\bibnamefont {Watanabe}}, \bibinfo {author}
  {\bibfnamefont {T.}~\bibnamefont {Taniguchi}}, \ and\ \bibinfo {author}
  {\bibfnamefont {P.}~\bibnamefont {Jarillo-Herrero}},\ }\href@noop {}
  {\bibfield  {journal} {\bibinfo  {journal} {arXiv preprint arXiv:2012.01434}\
  } (\bibinfo {year} {2020})}\BibitemShut {NoStop}%
\bibitem [{\citenamefont {Wang}\ \emph {et~al.}(2020)\citenamefont {Wang},
  \citenamefont {Zheng}, \citenamefont {Millis},\ and\ \citenamefont
  {Cano}}]{CanoChiral}%
  \BibitemOpen
  \bibfield  {author} {\bibinfo {author} {\bibfnamefont {J.}~\bibnamefont
  {Wang}}, \bibinfo {author} {\bibfnamefont {Y.}~\bibnamefont {Zheng}},
  \bibinfo {author} {\bibfnamefont {A.~J.}\ \bibnamefont {Millis}}, \ and\
  \bibinfo {author} {\bibfnamefont {J.}~\bibnamefont {Cano}},\ }\href@noop {}
  {\bibfield  {journal} {\bibinfo  {journal} {arXiv preprint arXiv:2010.03589}\
  } (\bibinfo {year} {2020})}\BibitemShut {NoStop}%
\bibitem [{\citenamefont {Lee}\ \emph {et~al.}(2019)\citenamefont {Lee},
  \citenamefont {Khalaf}, \citenamefont {Liu}, \citenamefont {Liu},
  \citenamefont {Hao}, \citenamefont {Kim},\ and\ \citenamefont
  {Vishwanath}}]{TDBGTheory}%
  \BibitemOpen
  \bibfield  {author} {\bibinfo {author} {\bibfnamefont {J.~Y.}\ \bibnamefont
  {Lee}}, \bibinfo {author} {\bibfnamefont {E.}~\bibnamefont {Khalaf}},
  \bibinfo {author} {\bibfnamefont {S.}~\bibnamefont {Liu}}, \bibinfo {author}
  {\bibfnamefont {X.}~\bibnamefont {Liu}}, \bibinfo {author} {\bibfnamefont
  {Z.}~\bibnamefont {Hao}}, \bibinfo {author} {\bibfnamefont {P.}~\bibnamefont
  {Kim}}, \ and\ \bibinfo {author} {\bibfnamefont {A.}~\bibnamefont
  {Vishwanath}},\ }\href {\doibase 10.1038/s41467-019-12981-1} {\bibfield
  {journal} {\bibinfo  {journal} {Nature Communications}\ } (\bibinfo {year}
  {2019}),\ 10.1038/s41467-019-12981-1},\ \Eprint
  {http://arxiv.org/abs/1903.08685} {arXiv:1903.08685} \BibitemShut {NoStop}%
\bibitem [{\citenamefont {Aoki}\ \emph
  {et~al.}(2019{\natexlab{a}})\citenamefont {Aoki}, \citenamefont {Nakamura},
  \citenamefont {Honda}, \citenamefont {Li}, \citenamefont {Homma},
  \citenamefont {Shimizu}, \citenamefont {Sato}, \citenamefont {Knebel},
  \citenamefont {Brison}, \citenamefont {Pourret}, \citenamefont {Braithwaite},
  \citenamefont {Lapertot}, \citenamefont {Niu}, \citenamefont {Vališka},
  \citenamefont {Harima},\ and\ \citenamefont {Flouquet}}]{UTe}%
  \BibitemOpen
  \bibfield  {author} {\bibinfo {author} {\bibfnamefont {D.}~\bibnamefont
  {Aoki}}, \bibinfo {author} {\bibfnamefont {A.}~\bibnamefont {Nakamura}},
  \bibinfo {author} {\bibfnamefont {F.}~\bibnamefont {Honda}}, \bibinfo
  {author} {\bibfnamefont {D.}~\bibnamefont {Li}}, \bibinfo {author}
  {\bibfnamefont {Y.}~\bibnamefont {Homma}}, \bibinfo {author} {\bibfnamefont
  {Y.}~\bibnamefont {Shimizu}}, \bibinfo {author} {\bibfnamefont {Y.~J.}\
  \bibnamefont {Sato}}, \bibinfo {author} {\bibfnamefont {G.}~\bibnamefont
  {Knebel}}, \bibinfo {author} {\bibfnamefont {J.-P.}\ \bibnamefont {Brison}},
  \bibinfo {author} {\bibfnamefont {A.}~\bibnamefont {Pourret}}, \bibinfo
  {author} {\bibfnamefont {D.}~\bibnamefont {Braithwaite}}, \bibinfo {author}
  {\bibfnamefont {G.}~\bibnamefont {Lapertot}}, \bibinfo {author}
  {\bibfnamefont {Q.}~\bibnamefont {Niu}}, \bibinfo {author} {\bibfnamefont
  {M.}~\bibnamefont {Vališka}}, \bibinfo {author} {\bibfnamefont
  {H.}~\bibnamefont {Harima}}, \ and\ \bibinfo {author} {\bibfnamefont
  {J.}~\bibnamefont {Flouquet}},\ }\href {\doibase 10.7566/JPSJ.88.043702}
  {\bibfield  {journal} {\bibinfo  {journal} {Journal of the Physical Society
  of Japan}\ }\textbf {\bibinfo {volume} {88}},\ \bibinfo {pages} {043702}
  (\bibinfo {year} {2019}{\natexlab{a}})},\ \Eprint
  {http://arxiv.org/abs/https://doi.org/10.7566/JPSJ.88.043702}
  {https://doi.org/10.7566/JPSJ.88.043702} \BibitemShut {NoStop}%
\bibitem [{\citenamefont {Aoki}\ \emph
  {et~al.}(2019{\natexlab{b}})\citenamefont {Aoki}, \citenamefont {Ishida},\
  and\ \citenamefont {Flouquet}}]{FerroSC}%
  \BibitemOpen
  \bibfield  {author} {\bibinfo {author} {\bibfnamefont {D.}~\bibnamefont
  {Aoki}}, \bibinfo {author} {\bibfnamefont {K.}~\bibnamefont {Ishida}}, \ and\
  \bibinfo {author} {\bibfnamefont {J.}~\bibnamefont {Flouquet}},\ }\href
  {\doibase 10.7566/JPSJ.88.022001} {\bibfield  {journal} {\bibinfo  {journal}
  {Journal of the Physical Society of Japan}\ }\textbf {\bibinfo {volume}
  {88}},\ \bibinfo {pages} {022001} (\bibinfo {year} {2019}{\natexlab{b}})},\
  \Eprint {http://arxiv.org/abs/https://doi.org/10.7566/JPSJ.88.022001}
  {https://doi.org/10.7566/JPSJ.88.022001} \BibitemShut {NoStop}%
\bibitem [{\citenamefont {Seo}\ \emph {et~al.}(2015)\citenamefont {Seo},
  \citenamefont {Kang}, \citenamefont {Kwon},\ and\ \citenamefont
  {Shin}}]{HQVBEC}%
  \BibitemOpen
  \bibfield  {author} {\bibinfo {author} {\bibfnamefont {S.~W.}\ \bibnamefont
  {Seo}}, \bibinfo {author} {\bibfnamefont {S.}~\bibnamefont {Kang}}, \bibinfo
  {author} {\bibfnamefont {W.~J.}\ \bibnamefont {Kwon}}, \ and\ \bibinfo
  {author} {\bibfnamefont {Y.-i.}\ \bibnamefont {Shin}},\ }\href {\doibase
  10.1103/PhysRevLett.115.015301} {\bibfield  {journal} {\bibinfo  {journal}
  {Phys. Rev. Lett.}\ }\textbf {\bibinfo {volume} {115}},\ \bibinfo {pages}
  {015301} (\bibinfo {year} {2015})}\BibitemShut {NoStop}%
\bibitem [{\citenamefont {Mukerjee}\ \emph {et~al.}(2006)\citenamefont
  {Mukerjee}, \citenamefont {Xu},\ and\ \citenamefont {Moore}}]{Mukerjee2006}%
  \BibitemOpen
  \bibfield  {author} {\bibinfo {author} {\bibfnamefont {S.}~\bibnamefont
  {Mukerjee}}, \bibinfo {author} {\bibfnamefont {C.}~\bibnamefont {Xu}}, \ and\
  \bibinfo {author} {\bibfnamefont {J.~E.}\ \bibnamefont {Moore}},\ }\href
  {\doibase 10.1103/PhysRevLett.97.120406} {\bibfield  {journal} {\bibinfo
  {journal} {Phys. Rev. Lett.}\ }\textbf {\bibinfo {volume} {97}},\ \bibinfo
  {pages} {120406} (\bibinfo {year} {2006})}\BibitemShut {NoStop}%
\bibitem [{\citenamefont {Radzihovsky}\ and\ \citenamefont
  {Vishwanath}(2009)}]{Leo_09}%
  \BibitemOpen
  \bibfield  {author} {\bibinfo {author} {\bibfnamefont {L.}~\bibnamefont
  {Radzihovsky}}\ and\ \bibinfo {author} {\bibfnamefont {A.}~\bibnamefont
  {Vishwanath}},\ }\href {\doibase 10.1103/PhysRevLett.103.010404} {\bibfield
  {journal} {\bibinfo  {journal} {Phys. Rev. Lett.}\ }\textbf {\bibinfo
  {volume} {103}},\ \bibinfo {pages} {010404} (\bibinfo {year}
  {2009})}\BibitemShut {NoStop}%
\bibitem [{\citenamefont {Berg}\ \emph {et~al.}(2009)\citenamefont {Berg},
  \citenamefont {Fradkin},\ and\ \citenamefont {Kivelson}}]{Berg_2009}%
  \BibitemOpen
  \bibfield  {author} {\bibinfo {author} {\bibfnamefont {E.}~\bibnamefont
  {Berg}}, \bibinfo {author} {\bibfnamefont {E.}~\bibnamefont {Fradkin}}, \
  and\ \bibinfo {author} {\bibfnamefont {S.~A.}\ \bibnamefont {Kivelson}},\
  }\href {\doibase 10.1038/nphys1389} {\bibfield  {journal} {\bibinfo
  {journal} {Nature Physics}\ }\textbf {\bibinfo {volume} {5}},\ \bibinfo
  {pages} {830–833} (\bibinfo {year} {2009})}\BibitemShut {NoStop}%
\bibitem [{\citenamefont {Jian}\ and\ \citenamefont {Zhai}(2011)}]{Jian2011}%
  \BibitemOpen
  \bibfield  {author} {\bibinfo {author} {\bibfnamefont {C.-M.}\ \bibnamefont
  {Jian}}\ and\ \bibinfo {author} {\bibfnamefont {H.}~\bibnamefont {Zhai}},\
  }\href {\doibase 10.1103/PhysRevB.84.060508} {\bibfield  {journal} {\bibinfo
  {journal} {Phys. Rev. B}\ }\textbf {\bibinfo {volume} {84}},\ \bibinfo
  {pages} {060508} (\bibinfo {year} {2011})}\BibitemShut {NoStop}%
\bibitem [{\citenamefont {Zhang}\ \emph {et~al.}(2019)\citenamefont {Zhang},
  \citenamefont {Mao}, \citenamefont {Cao}, \citenamefont {Jarillo-Herrero},\
  and\ \citenamefont {Senthil}}]{Zhang2018}%
  \BibitemOpen
  \bibfield  {author} {\bibinfo {author} {\bibfnamefont {Y.-H.}\ \bibnamefont
  {Zhang}}, \bibinfo {author} {\bibfnamefont {D.}~\bibnamefont {Mao}}, \bibinfo
  {author} {\bibfnamefont {Y.}~\bibnamefont {Cao}}, \bibinfo {author}
  {\bibfnamefont {P.}~\bibnamefont {Jarillo-Herrero}}, \ and\ \bibinfo {author}
  {\bibfnamefont {T.}~\bibnamefont {Senthil}},\ }\href {\doibase
  10.1103/PhysRevB.99.075127} {\bibfield  {journal} {\bibinfo  {journal} {Phys.
  Rev. B}\ }\textbf {\bibinfo {volume} {99}},\ \bibinfo {pages} {075127}
  (\bibinfo {year} {2019})}\BibitemShut {NoStop}%
\bibitem [{\citenamefont {Khokhlachev}(1976)}]{Khokhlachev1976}%
  \BibitemOpen
  \bibfield  {author} {\bibinfo {author} {\bibfnamefont {S.}~\bibnamefont
  {Khokhlachev}},\ }\href@noop {} {\bibfield  {journal} {\bibinfo  {journal}
  {Sov. Phys. JETP}\ }\textbf {\bibinfo {volume} {43}},\ \bibinfo {pages} {137}
  (\bibinfo {year} {1976})}\BibitemShut {NoStop}%
\bibitem [{\citenamefont {R.A.~Pelcovits}(1976)}]{Pelcovits1976}%
  \BibitemOpen
  \bibfield  {author} {\bibinfo {author} {\bibfnamefont {D.~N.}\ \bibnamefont
  {R.A.~Pelcovits}},\ }\href@noop {} {\bibfield  {journal} {\bibinfo  {journal}
  {Physics Letters A}\ }\textbf {\bibinfo {volume} {57}},\ \bibinfo {pages}
  {23} (\bibinfo {year} {1976})}\BibitemShut {NoStop}%
\bibitem [{\citenamefont {Hikami}\ and\ \citenamefont
  {Tsuneto}(1980)}]{Hikami1980}%
  \BibitemOpen
  \bibfield  {author} {\bibinfo {author} {\bibfnamefont {S.}~\bibnamefont
  {Hikami}}\ and\ \bibinfo {author} {\bibfnamefont {T.}~\bibnamefont
  {Tsuneto}},\ }\href {\doibase 10.1143/PTP.63.387} {\bibfield  {journal}
  {\bibinfo  {journal} {Progress of Theoretical Physics}\ }\textbf {\bibinfo
  {volume} {63}},\ \bibinfo {pages} {387} (\bibinfo {year} {1980})},\ \Eprint
  {http://arxiv.org/abs/https://academic.oup.com/ptp/article-pdf/63/2/387/5332385/63-2-387.pdf}
  {https://academic.oup.com/ptp/article-pdf/63/2/387/5332385/63-2-387.pdf}
  \BibitemShut {NoStop}%
\bibitem [{\citenamefont {Cuccoli}\ \emph {et~al.}(1995)\citenamefont
  {Cuccoli}, \citenamefont {Tognetti},\ and\ \citenamefont
  {Vaia}}]{Cuccoli1995}%
  \BibitemOpen
  \bibfield  {author} {\bibinfo {author} {\bibfnamefont {A.}~\bibnamefont
  {Cuccoli}}, \bibinfo {author} {\bibfnamefont {V.}~\bibnamefont {Tognetti}}, \
  and\ \bibinfo {author} {\bibfnamefont {R.}~\bibnamefont {Vaia}},\ }\href
  {\doibase 10.1103/PhysRevB.52.10221} {\bibfield  {journal} {\bibinfo
  {journal} {Phys. Rev. B}\ }\textbf {\bibinfo {volume} {52}},\ \bibinfo
  {pages} {10221} (\bibinfo {year} {1995})}\BibitemShut {NoStop}%
\bibitem [{\citenamefont {Cornfeld}\ \emph {et~al.}(2020)\citenamefont
  {Cornfeld}, \citenamefont {Rudner},\ and\ \citenamefont
  {Berg}}]{cornfeld2020spinpolarized}%
  \BibitemOpen
  \bibfield  {author} {\bibinfo {author} {\bibfnamefont {E.}~\bibnamefont
  {Cornfeld}}, \bibinfo {author} {\bibfnamefont {M.~S.}\ \bibnamefont
  {Rudner}}, \ and\ \bibinfo {author} {\bibfnamefont {E.}~\bibnamefont
  {Berg}},\ }\href@noop {} {\enquote {\bibinfo {title} {Spin-polarized
  superconductivity: order parameter topology, current dissipation, and
  multiple-period josephson effect},}\ } (\bibinfo {year} {2020}),\ \Eprint
  {http://arxiv.org/abs/2006.10073} {arXiv:2006.10073 [cond-mat.supr-con]}
  \BibitemShut {NoStop}%
\bibitem [{\citenamefont {Khalaf}\ and\ \citenamefont
  {Vishwanath}()}]{SkParton}%
  \BibitemOpen
  \bibfield  {author} {\bibinfo {author} {\bibfnamefont {E.}~\bibnamefont
  {Khalaf}}\ and\ \bibinfo {author} {\bibfnamefont {A.}~\bibnamefont
  {Vishwanath}},\ }\href@noop {} {\ }\bibinfo {note} {{i}n
  preparation}\BibitemShut {NoStop}%
\bibitem [{\citenamefont {Moon}\ \emph {et~al.}(1995)\citenamefont {Moon},
  \citenamefont {Mori}, \citenamefont {Yang}, \citenamefont {Girvin},
  \citenamefont {MacDonald}, \citenamefont {Zheng}, \citenamefont {Yoshioka},\
  and\ \citenamefont {Zhang}}]{MoonMori}%
  \BibitemOpen
  \bibfield  {author} {\bibinfo {author} {\bibfnamefont {K.}~\bibnamefont
  {Moon}}, \bibinfo {author} {\bibfnamefont {H.}~\bibnamefont {Mori}}, \bibinfo
  {author} {\bibfnamefont {K.}~\bibnamefont {Yang}}, \bibinfo {author}
  {\bibfnamefont {S.~M.}\ \bibnamefont {Girvin}}, \bibinfo {author}
  {\bibfnamefont {A.~H.}\ \bibnamefont {MacDonald}}, \bibinfo {author}
  {\bibfnamefont {L.}~\bibnamefont {Zheng}}, \bibinfo {author} {\bibfnamefont
  {D.}~\bibnamefont {Yoshioka}}, \ and\ \bibinfo {author} {\bibfnamefont
  {S.-C.}\ \bibnamefont {Zhang}},\ }\href {\doibase 10.1103/PhysRevB.51.5138}
  {\bibfield  {journal} {\bibinfo  {journal} {Phys. Rev. B}\ }\textbf {\bibinfo
  {volume} {51}},\ \bibinfo {pages} {5138} (\bibinfo {year}
  {1995})}\BibitemShut {NoStop}%
\bibitem [{\citenamefont {{Cao}}\ \emph {et~al.}(2020)\citenamefont {{Cao}},
  \citenamefont {{Rodan-Legrain}}, \citenamefont {{Park}}, \citenamefont {{Noah
  Yuan}}, \citenamefont {{Watanabe}}, \citenamefont {{Taniguchi}},
  \citenamefont {{Fernandes}}, \citenamefont {{Fu}},\ and\ \citenamefont
  {{Jarillo-Herrero}}}]{PabloNematic}%
  \BibitemOpen
  \bibfield  {author} {\bibinfo {author} {\bibfnamefont {Y.}~\bibnamefont
  {{Cao}}}, \bibinfo {author} {\bibfnamefont {D.}~\bibnamefont
  {{Rodan-Legrain}}}, \bibinfo {author} {\bibfnamefont {J.~M.}\ \bibnamefont
  {{Park}}}, \bibinfo {author} {\bibfnamefont {F.}~\bibnamefont {{Noah Yuan}}},
  \bibinfo {author} {\bibfnamefont {K.}~\bibnamefont {{Watanabe}}}, \bibinfo
  {author} {\bibfnamefont {T.}~\bibnamefont {{Taniguchi}}}, \bibinfo {author}
  {\bibfnamefont {R.~M.}\ \bibnamefont {{Fernandes}}}, \bibinfo {author}
  {\bibfnamefont {L.}~\bibnamefont {{Fu}}}, \ and\ \bibinfo {author}
  {\bibfnamefont {P.}~\bibnamefont {{Jarillo-Herrero}}},\ }\href@noop {}
  {\bibfield  {journal} {\bibinfo  {journal} {arXiv e-prints}\ ,\ \bibinfo
  {pages} {2004.04148}} (\bibinfo {year} {2020})}\BibitemShut {NoStop}%
\bibitem [{\citenamefont {Ran}\ \emph {et~al.}(2019{\natexlab{b}})\citenamefont
  {Ran}, \citenamefont {Eckberg}, \citenamefont {Ding}, \citenamefont
  {Furukawa}, \citenamefont {Metz}, \citenamefont {Saha}, \citenamefont {Liu},
  \citenamefont {Zic}, \citenamefont {Kim}, \citenamefont {Paglione},\ and\
  \citenamefont {Butch}}]{Ran2019}%
  \BibitemOpen
  \bibfield  {author} {\bibinfo {author} {\bibfnamefont {S.}~\bibnamefont
  {Ran}}, \bibinfo {author} {\bibfnamefont {C.}~\bibnamefont {Eckberg}},
  \bibinfo {author} {\bibfnamefont {Q.-P.}\ \bibnamefont {Ding}}, \bibinfo
  {author} {\bibfnamefont {Y.}~\bibnamefont {Furukawa}}, \bibinfo {author}
  {\bibfnamefont {T.}~\bibnamefont {Metz}}, \bibinfo {author} {\bibfnamefont
  {S.~R.}\ \bibnamefont {Saha}}, \bibinfo {author} {\bibfnamefont {I.-L.}\
  \bibnamefont {Liu}}, \bibinfo {author} {\bibfnamefont {M.}~\bibnamefont
  {Zic}}, \bibinfo {author} {\bibfnamefont {H.}~\bibnamefont {Kim}}, \bibinfo
  {author} {\bibfnamefont {J.}~\bibnamefont {Paglione}}, \ and\ \bibinfo
  {author} {\bibfnamefont {N.~P.}\ \bibnamefont {Butch}},\ }\href {\doibase
  10.1126/science.aav8645} {\bibfield  {journal} {\bibinfo  {journal}
  {Science}\ }\textbf {\bibinfo {volume} {365}},\ \bibinfo {pages} {684}
  (\bibinfo {year} {2019}{\natexlab{b}})},\ \Eprint
  {http://arxiv.org/abs/https://science.sciencemag.org/content/365/6454/684.full.pdf}
  {https://science.sciencemag.org/content/365/6454/684.full.pdf} \BibitemShut
  {NoStop}%
\bibitem [{\citenamefont {Nevidomskyy}(2020{\natexlab{b}})}]{nevidomskyy2020}%
  \BibitemOpen
  \bibfield  {author} {\bibinfo {author} {\bibfnamefont {A.~H.}\ \bibnamefont
  {Nevidomskyy}},\ }\href@noop {} {\enquote {\bibinfo {title} {Stability of a
  nonunitary triplet pairing on the border of magnetism in ute$_2$},}\ }
  (\bibinfo {year} {2020}{\natexlab{b}}),\ \Eprint
  {http://arxiv.org/abs/2001.02699} {arXiv:2001.02699 [cond-mat.supr-con]}
  \BibitemShut {NoStop}%
\bibitem [{\citenamefont {Jang}\ \emph {et~al.}(2011)\citenamefont {Jang},
  \citenamefont {Ferguson}, \citenamefont {Vakaryuk}, \citenamefont {Budakian},
  \citenamefont {Chung}, \citenamefont {Goldbart},\ and\ \citenamefont
  {Maeno}}]{Jang2011}%
  \BibitemOpen
  \bibfield  {author} {\bibinfo {author} {\bibfnamefont {J.}~\bibnamefont
  {Jang}}, \bibinfo {author} {\bibfnamefont {D.~G.}\ \bibnamefont {Ferguson}},
  \bibinfo {author} {\bibfnamefont {V.}~\bibnamefont {Vakaryuk}}, \bibinfo
  {author} {\bibfnamefont {R.}~\bibnamefont {Budakian}}, \bibinfo {author}
  {\bibfnamefont {S.~B.}\ \bibnamefont {Chung}}, \bibinfo {author}
  {\bibfnamefont {P.~M.}\ \bibnamefont {Goldbart}}, \ and\ \bibinfo {author}
  {\bibfnamefont {Y.}~\bibnamefont {Maeno}},\ }\href {\doibase
  10.1126/science.1193839} {\bibfield  {journal} {\bibinfo  {journal}
  {Science}\ }\textbf {\bibinfo {volume} {331}},\ \bibinfo {pages} {186}
  (\bibinfo {year} {2011})},\ \Eprint
  {http://arxiv.org/abs/https://science.sciencemag.org/content/331/6014/186.full.pdf}
  {https://science.sciencemag.org/content/331/6014/186.full.pdf} \BibitemShut
  {NoStop}%
\bibitem [{\citenamefont {Volovik}(1999)}]{Volovik1999}%
  \BibitemOpen
  \bibfield  {author} {\bibinfo {author} {\bibfnamefont {G.~E.}\ \bibnamefont
  {Volovik}},\ }\href {\doibase 10.1134/1.568223} {\bibfield  {journal}
  {\bibinfo  {journal} {Journal of Experimental and Theoretical Physics
  Letters}\ }\textbf {\bibinfo {volume} {70}},\ \bibinfo {pages} {609}
  (\bibinfo {year} {1999})}\BibitemShut {NoStop}%
\bibitem [{\citenamefont {Read}\ and\ \citenamefont {Green}(2000)}]{Read2000}%
  \BibitemOpen
  \bibfield  {author} {\bibinfo {author} {\bibfnamefont {N.}~\bibnamefont
  {Read}}\ and\ \bibinfo {author} {\bibfnamefont {D.}~\bibnamefont {Green}},\
  }\href {\doibase 10.1103/PhysRevB.61.10267} {\bibfield  {journal} {\bibinfo
  {journal} {Phys. Rev. B}\ }\textbf {\bibinfo {volume} {61}},\ \bibinfo
  {pages} {10267} (\bibinfo {year} {2000})}\BibitemShut {NoStop}%
\bibitem [{\citenamefont {Ivanov}(2001)}]{Ivanov2001}%
  \BibitemOpen
  \bibfield  {author} {\bibinfo {author} {\bibfnamefont {D.~A.}\ \bibnamefont
  {Ivanov}},\ }\href {\doibase 10.1103/PhysRevLett.86.268} {\bibfield
  {journal} {\bibinfo  {journal} {Phys. Rev. Lett.}\ }\textbf {\bibinfo
  {volume} {86}},\ \bibinfo {pages} {268} (\bibinfo {year} {2001})}\BibitemShut
  {NoStop}%
\end{thebibliography}%
 
\pagebreak
\widetext
\begin{center}

\textbf{\large Supplemental material:\\ Symmetry constraints on superconducting pairing in twisted bilayer graphene}
\end{center}
\setcounter{equation}{0}
\setcounter{figure}{0}
\setcounter{table}{0}
\setcounter{page}{1}
\makeatletter
\renewcommand{\theequation}{S\arabic{equation}}
\renewcommand{\thefigure}{S\arabic{figure}}
\renewcommand{\bibnumfmt}[1]{[S#1]}
\setcounter{section}{0}
 
 \section{Transformation under spatial symmetries}
 In the main text, we only considered the transformation properties of the pairing function under global symmetries. Here, we consider the spatial symmetries. We follow the notations of Refs.~\cite{KIVCpaper, SkPaper} by writing the microscopic symmetries acting on the microscopic annihilation operators $f_\bk$ which is a vector in valley ($\tau$), sublattice ($\sigma$), spin ($s$), and layer ($\mu$):
\beq
C_3: f_\bk \mapsto  e^{i\frac{\pi}{3} \sigma_z \tau_z} f_{O_3 \bk}, \qquad C_2: f_\bk \mapsto \sigma_x \tau_x f_{-\bk}, \qquad M_y: f_\bk\mapsto \sigma_x \mu_x f_{m_y \bk}
\eeq
where $O_3$ denotes the standard action of threefold rotation on a 2D vector and $m_y (k_x, k_y) = (k_x, -k_y)$. Note that all these symmetries acts trivially on the spin index due to the $\SU(2)$ rotation symmetry. Next, we project these symmetries onto the flat bands. Due to the finite sublattice polarization, we can label the wavefunctions with the sublattice $\sigma$ their weight is concentrated on \cite{KIVCpaper}, which means that the flatband electron annihilation operators are labelled by valley ($\tau$), sublattice $(\sigma)$, and spin ($s$). A more convenient basis is obtained by replacing the valley-sublattice indices with the the Chern $(\gamma)$ - pseudospin $(\eta)$ indices defined via \cite{SkPaper}:
\beq
{\boldsymbol \gamma} = (\sigma_x, \sigma_y \tau_z, \sigma_z \tau_z), \qquad {\boldsymbol \eta} = (\sigma_x \tau_x, \sigma_x \tau_y, \tau_z)
\eeq
Due to the non-trivial band topology of the flat bands, some symmetries acquire a $\bk$-dependence when projected onto them. This dependence was derived in detail in Refs.~\cite{KIVCpaper, SkPaper} leading to the results summarized in the table below
\begin{table}[h]
\begin{tabular}{c|c|c|c}
\hline \hline
basis & $C_2$ & $C_3$ & $M_y$ \\
\hline
microscopic & $\sigma_x \tau_x$ & $e^{i \frac{\pi}{3} \sigma_z \tau_z}$ & $\sigma_x \mu_x$\\
\hline
projected & $\sigma_x \tau_x e^{i \theta_2(\bk)}$ & $e^{i \theta_3(\bk) \sigma_z \tau_z}$ & $\sigma_x e^{i \theta_y(\bk) \sigma_z \tau_z}$ \\
\hline
projected $(\gamma,\eta)$ & $\eta_x e^{i \theta_2(\bk)}$ & $e^{i \theta_3(\bk) \gamma_z}$ & $\gamma_x e^{i \theta_y(\bk) \gamma_z}$\\
\hline \hline
\end{tabular}
\end{table}

Let us now consider the transformation properties of the pairing function. We restrict ourselves to the physically relevant case of opposite momentum pairing:
\beq
\Delta_\bk = c_\bk c_{-\bk}^T
\eeq
which behaves under spatial symmetries as
\beq
c_\bk \mapsto U_\bk c_{O \bk} \qquad \implies \qquad \Delta_\bk \mapsto U_\bk \Delta_{O \bk} U_{-\bk}^T
\eeq
As in the main text (Eq.~3), we can use the fact the global symmetries do not have structure in the Chern space $\gamma$ to decompose the gap function into different channels
\beq
\Delta(\bk) = \Delta_x(\bk) \gamma_x + \Delta_y(\bk) \gamma_y + \Delta_+(\bk) \frac{1 + \gamma_z}{2} + \Delta_-(\bk) \frac{1 - \gamma_z}{2}
\eeq
The antisymmetry of the full pairing function $\Delta(\bk)^T = -\Delta(-\bk)$ means that
\beq
\Delta_{x,\pm}(\bk)^T = -\Delta_{x,\pm}(-\bk), \qquad \Delta_y(\bk)^T = \Delta_y(-\bk)
\label{Deltaxpm}
\eeq
The transformation properties under $C_2$, $C_3$ and $M_y$ can now be obtained:
\begin{gather}
C_2: \Delta_{x,y,\pm}(\bk) \mapsto \eta_x \Delta_{x,y,\pm}(-\bk) \eta_x, \\
C_3: \Delta_{x,y}(\bk) \mapsto  \Delta_{x,y}(O_3\bk), \quad  \Delta_\pm(\bk) \mapsto e^{\pm 2 i \theta_3(\bk)} \Delta_\pm(O_3\bk) \\
M_y: \Delta_{x}(\bk) \mapsto \Delta_x(m_y \bk), \quad \Delta_{y}(\bk) \mapsto -\Delta_y(m_y \bk), \quad  \Delta_\pm(\bk) \mapsto e^{\pm 2 i \theta_y(\bk)}\Delta_\mp(m_y \bk).
\end{gather}
The last two equations imply that the $\Delta_\pm$ pairing channels are tied together in a single irrep of the spatial symmetry group, whereas $\Delta_x$ and $\Delta_y$ transform as separate irreps. To understand the action of $C_2$, we decompose each $\Delta_\mu$ into symmetric and antisymmetric intervalley pairing and an intravalley pairing. For intervalley pairing considered in the main text
\beq
\Delta^{\rm S/A}(\bk) = \left( \begin{array}{cc}
    0 & \tilde \Delta^{\rm S/A}(\bk) \\
    \pm \tilde \Delta^{\rm S/A}(\bk)^T & 0
\end{array} \right) \quad \implies \quad \tilde \Delta^{\rm S/A}(\bk) = \pm \tilde \Delta^{\rm S/A}(-\bk)^T 
\eeq
Combining with Eq.~\ref{Deltaxpm} yields
\beq
\tilde \Delta^{\rm S/A}_{x,\pm}(\bk)^T = \mp \tilde \Delta^{\rm S/A}_{x,\pm}(\bk), \quad \tilde \Delta^{\rm S/A}_y(\bk)^T = \pm \tilde \Delta^{\rm S/A}_y(\bk)
\eeq
This represents a restriction on the individual blocks $\tilde \Delta$ in the different pairing channels $\pm, x, y$ for symmetric and antisymmetric pairing functions. Finally, we consider intravalley pairing
\beq
\Delta(\bk) = \left( \begin{array}{cc}
    \Delta_K(\bk) & 0 \\
    0 & \Delta_{K'}(\bk)
\end{array} \right), \quad \implies \quad \Delta_K(\bk) = \Delta_{K'}(-\bk)
\label{DeltaIntra}
\eeq
which dictates that the pairing in both valleys has the same gap and symmetry properties (singlet or triplet).
 
 \section{Fractional vortices for intravalley pairing}
Let us now consider the case of intra-valley pairing. This is not likely to be the relevant case physically but we include it here for completeness. Assuming unitary pairing, we can write the most general form for intra-valley pairing as
\beq
\Delta = \left(\begin{array}{cc} \Delta_K & 0 \\ 0 & \Delta_{K'} \end{array} \right), \qquad \Delta_K \Delta_{K}^\dagger = \Delta_{K'} \Delta_{K'}^\dagger \propto \mathbbm{1}
\eeq
If both $\Delta_K$ and $\Delta_{K'}$ are antisymmetric, which corresponds to the singlet channel of $\SU(2)$ in each valley, then we can use the same considerations as in the main text to determine the type of vortices with the GL Lagrangian
\beq
 \L = \rho_K (\partial_\mu \varphi_K - \frac{2 e}{\hbar} A_\mu)^2 + \rho_{K'} (\partial_\mu \varphi_{K'} - \frac{2 e}{\hbar} A_\mu)^2 + \dots
 \eeq
 leading to the vortex quantization condition
 \beq
\int d l_\mu A_\mu = \frac{ \hbar}{2 e} \int d l_\mu \frac{\rho_{K} \partial_\mu \varphi_{K} + \rho_{K'} \partial_\mu \varphi_{K'}}{\rho_K + \rho_{K'}}
\label{FluxKK}
 \eeq
 Symmetries such as $C_2$ and time reversal exchange valleys and therefore relate $\Delta_K$ and $\Delta_{K'}$ as well as $\rho_K$ and $\rho_{K'}$.  As long as the superconductivity develops in a state that preserves one of $C_2$, time reversal or any combination of one of them with an internal symmetry, we expect $\rho_K = \rho_{K'}$ for unitary pairing. We then obtain $h/4e$ vortices.
 
 This conclusions can also be verified by considering the manifold of the superconducting order parameter by writing $\Delta = s_y e^{i (\phi \eta_0 + \phi_V \eta_z)}$. Thus, we can form a loop in the space of $\Delta$ where both $\phi$ and $\phi_V$ go from 0 to $\pi$. This will correspond to a half-vortex in the overall superconducting phase $\phi$. This reflects the nature of the manifold of the superconducting order parameter $\frac{\U_C(1) \times \U_V(1)}{\Z_2}$ which admits half-vortices. This case is similar to the case of non-unitary pairing considered in the main text where $\pi_1$ of the superconducting manifold is $\Z \oplus \Z$ and vortices with arbitrary flux $x h/2e$ are in principle allowed. However, the presence of any symmetry relating the two valleys would force $x = 1/2$ leading to half quantum vortices.
 
 If both $C_2$ and time reversal are broken in the parent state and there is no other unbroken symmetry that relates the two valleys then $\rho_K$ and $\rho_{K'}$ may be different. In such a state we would expect nonunitary pairing since there is also no symmetry relating $\Delta_K$ and $\Delta_{K'}$ which means unitary pairing should not be stable to RG flows; it must be fine tuned.  If we nonetheless fine tune to obtain unitary pairing, and furthermore fine tune one of the stiffnesses to zero despite a nonzero pairing potential, we obtain conventional $h/2e$ vortices.  We do not take this latter scenario into account for our main theorem due to the extreme amounts of unphysical fine tuning, but detailed its existence here for completeness.

 Next, we consider the case where either $\Delta_K$ or $\Delta_{K'}$ is symmetric, i.e.  $\SU(2)$ triplet. The existence of fractional vortices in this case follows immediately from the existence of half-vortices in the spin-triplet case whose manifold is $\U(2)/\O(2)$ with $\pi_1(\U(2)/\O(2)) = \Z/2$. When included in the bigger order parameter for both valleys, this yields fractional vortices.  For instance, if pairing is triplet in both valleys, this will be a quarter vortex similar to the symmetric $\SU(4)$ case considered in the main text.
 
 \end{document}